\documentstyle[12pt]{article}
\def\lsim{\raisebox{-4pt}{$\,\stackrel{\textstyle{<}}{\sim}\,$}}
\def\gsim{\raisebox{-4pt}{$\,\stackrel{\textstyle{>}}{\sim}\,$}}
\begin{document}
\begin{flushright}
\baselineskip=12pt
DOE/ER/40717--36\\
CTP-TAMU-58/96\\
ACT-17/96\\
\tt hep-ph/9611437
\end{flushright}

\begin{center}
\vglue 1.0cm
{\Large\bf Single-photon signals at LEP in supersymmetric models with a light gravitino}
\vglue 1.0cm
{\Large Jorge L. Lopez$^1$, D.V. Nanopoulos$^{2,3}$, and A.~Zichichi$^4$}
\vglue 1cm
\begin{flushleft}
$^1$ Bonner Nuclear Lab, Department of Physics, Rice University\\ 6100 Main
Street, Houston, TX 77005, USA\\
$^2$Center for Theoretical Physics, Department of Physics, Texas A\&M
University\\ College Station, TX 77843--4242, USA\\
$^3$Astroparticle Physics Group, Houston Advanced Research Center (HARC)\\
The Mitchell Campus, The Woodlands, TX 77381, USA\\
$^4$University and INFN--Bologna, Italy and CERN, 1211 Geneva 23, Switzerland\\
\end{flushleft}
\end{center}

\vglue 1cm
\begin{abstract}
We study the single-photon signals expected at LEP in models with a very light gravitino. The dominant process is neutralino-gravitino production ($e^+e^-\to \chi\widetilde G$) with subsequent neutralino decay via $\chi\to\gamma\widetilde G$, giving a $\gamma+{\rm E_{miss}}$ signal. We first 
calculate the cross section at arbitrary center-of-mass energies and provide new analytic expressions for the differential cross section valid for general
neutralino compositions. We then consider the constraints on the gravitino mass from LEP~1 and LEP161 single-photon searches, and possible such searches at
the Tevatron. We show that it is possible to evade the stringent LEP~1 limits and still obtain an observable rate at LEP~2, in particular in the region of parameter space that may explain the CDF $ee\gamma\gamma+{\rm E_{T,miss}}$ event. As diphoton events from neutralino pair-production would not be kinematically accessible in this scenario, the observation of whichever photonic signal will discriminate among the various light-gravitino scenarios in the literature. We also perform a Monte Carlo simulation of the expected energy and angular distributions of the emitted photon, and of the missing invariant mass expected in the events. Finally we specialize the results to the case of a recently proposed one-parameter no-scale supergravity model.
\end{abstract}

\vspace{0.5cm}
\begin{flushleft}
\baselineskip=12pt
November 1996\\
\end{flushleft}
\newpage
\setcounter{page}{1}
\pagestyle{plain}
\baselineskip=14pt

\section{Introduction}
\label{sec:introduction}
Supersymmetric models with a light gravitino ($\widetilde G$) have been considered for some time \cite{FayetEarly,EEN,Fayet,Dicus,DicusZ}, but interest on them has recently surged \cite{DineKane,Gravitino} because of their ability to explain naturally the puzzling $e^+e^-\gamma\gamma+{\rm E_{T,miss}}$ event observed by the CDF Collaboration \cite{Park}. If the gravitino is the lightest supersymmetric particle (LSP),\footnote{We assume that R-parity is conserved,
as otherwise the decay $p\to\widetilde G K^+$ may occur at an unsuppressed
rate \cite{CCL}.} the next-to-lightest supersymmetric particle (NLSP, typically
the lightest neutralino ($\chi$), as we will assume here) becomes unstable and eventually decays into a photon plus a gravitino 
($\chi\to\gamma\widetilde G$) \cite{EEN}. This decay becomes of experimental interest when it happens quickly enough for the photon to be observed in the detector. Because the interaction of the gravitino with matter is inversely proportional to the gravitino mass, the neutralino lifetime will be short enough for a sufficiently light gravitino: $m_{\tilde G}\lsim 250\,{\rm eV}$ \cite{DineKane}. On the other hand, the gravitino may not be too light, as otherwise it would be copiously produced leading to distinctive signals at colliders that have not been observed \cite{Fayet,DicusZ} or cosmological \cite{MoroiGherghetta} and astrophysical \cite{astro} embarrassments:
$m_{\tilde G}>10^{-6}\,{\rm eV}$. LEP~1 searches strengthen this limit to $m_{\tilde G}\gsim10^{-3}\,{\rm eV}$ in large regions of parameter space, when $m_\chi<M_Z$ \cite{1vs2gamma}.

Theoretically, light gravitinos are expected in gauge-mediated models of
low-energy supersymmetry \cite{DineKane}, where the gravitino mass is related to the scale of supersymmetry breaking via
$m_{\tilde G}\approx6\times10^{-5}\,{\rm eV}\,(\Lambda_{\rm SUSY}/500
\,{\rm GeV})^2$. Special cases of gravity-mediated models may also yield light
gravitinos, when the scale of local and global breaking of supersymmetry
are decoupled, as in the context of no-scale supergravity \cite{EEN,Gravitino},
in which case $m_{\tilde G}\sim (m_{1/2}/M_{Pl})^p\, M_{Pl}$, with $m_{1/2}$
the gaugino mass scale and $p\sim2$ a model-dependent constant. 

Experimental searches for supersymmetry are considerably more sensitive in this type of neutralino-unstable supersymmetric models. First of all, the lightest observable supersymmetric channel is no longer a pair of charginos ($\chi^+\chi^-$), but instead a pair of (the usually lighter) neutralinos ($\chi\chi$), or if the gravitino is light enough ($m_{\tilde G}\lsim 10^{-3}\,{\rm eV}$) the neutralino-gravitino channel ($\chi\widetilde G$). These new channels allow a deeper exploration into parameter space. Furthermore, because of the photonic signature in all supersymmetric processes, it becomes possible to overcome the loss of experimental sensitivity that occurs when the daughter leptons become too soft (as in chargino pair production when $m_{\chi^\pm}-m_\chi<10\,{\rm GeV}$ or $m_{\chi^\pm}>m_{\tilde\nu}>m_{\chi^\pm}-3\,{\rm GeV}$), and therefore absolute lower bounds on sparticle masses become experimentally attainable in this class of models. Indeed, diphoton searches at LEP161 ({\em i.e.}, $\sqrt{s}=161\,{\rm GeV}$) \cite{Oct8} have been recently shown \cite{October8} to exclude a significant fraction of the parameter space that is preferred by the supersymmetric interpretations of the CDF event.\footnote{We should emphasize that even though we are encouraged by the natural interpretation of the CDF event within certain light gravitino scenarios, we believe that such scenarios are interesting in their own right and should be fully explored irrespective of the status of the CDF event.} Ongoing runs at LEP~2 should be able to probe even deeper into the remaining preferred region of parameter space. 

Our purpose here is to consider in detail a complementary signal in 
light-gravitino models, namely the associated production of gravitinos with
neutralinos.\footnote{Gravitino pair production ($e^+e^-\to\widetilde G\widetilde G$) exceeds neutralino-gravitino production ($e^+e^-\to\chi\widetilde G$) only for gravitino masses that have already
been excluded experimentally ($m_{\tilde G}\lsim10^{-6}\,{\rm eV}$). The
single-photon signal is further suppressed by a factor of $\alpha$ from
the radiated photon.} The resulting single-photon signal has been recently shown to be observable at LEP~2 in certain range of gravitino masses, but only when the diphoton signal from neutralino pair production is itself not kinematically accessible \cite{1vs2gamma}. Therefore, experimental observation of whichever photonic signal will provide very useful information in sorting out the various light-gravitino scenarios in the literature. The gravitino mass plays a central role in gravitino-production processes, whose rate is inversely proportional to the gravitino mass squared ($1/m^2_{\tilde G}$). In contrast, the precise value of the gravitino mass plays a minor role in the production of the traditional supersymmetric particles, as it determines only the decay length of the neutralino. The neutralino-gravitino process of interest at LEP is
\begin{equation}
\sigma(e^+e^-\to\chi\widetilde G\to\gamma+{\rm E_{miss}})
\propto {\beta^8\over m^2_{\widetilde G}}\,,\quad{\rm with}\quad \beta=
\sqrt{1-{m^2_\chi\over s}}\ ,
\label{eq:process}
\end{equation}
which provides an experimental handle on the gravitino mass. This process was considered originally by Fayet \cite{Fayet} (in the restricted case of a very light photino-like neutralino) who noted that the $\beta^8$ threshold behavior in Eq.~(\ref{eq:process}) results from subtle cancellations among all
contributing amplitudes. Dimensional analysis indicates that this cross section exceeds electroweak strength when $M^4_Z/(M^2_{\rm Pl}\,m^2_{\widetilde G})\gsim\alpha_{\rm weak}$ or $m_{\widetilde G}\lsim \alpha^{-1/2}_{\rm weak}\,
M^2_Z/M_{\rm Pl}\sim10^{-4}\,{\rm eV}$. In the context of LEP~1, this process was revisited in the restricted case of a neutralino with a
non-negligible zino component, where the resonant $Z$-exchange diagram dominates \cite{DicusZ}. 

In this paper we first calculate the cross section for the $e^+e^-\to\chi\widetilde G$ process at arbitrary center-of-mass energies and give new analytic expressions for the corresponding differential cross section (Sec.~\ref{sec:cross}). Next we reassess the constraints on the gravitino mass in view of the full LEP~1 data set and imposing the preliminary limits obtained recently from runs at LEP161 ($\sqrt{s}=161\,{\rm GeV}$), for general neutralino compositions (Sec.~\ref{sec:limits}). We also comment on the potential of analogous searches at the Tevatron.
We then perform a Monte Carlo simulation of the production and decay processes leading to the single-photon signal and obtain energy ($E_\gamma$) and angular ($\cos\theta_\gamma$) distributions for representative points in parameter space (Sec.~\ref{sec:signal}), and also discuss the missing invariant mass
distribution expected in the events.
We show that one may evade the LEP~1 limits and still obtain observable single-photon signals at LEP~2, although only when the diphoton signal from neutralino pair production is kinematically inaccessible. Finally we specialize our results to the case of our proposed one-parameter no-scale supergravity model \cite{Gravitino,One} (Sec.~\ref{sec:models}). Our conclusions are summarized
in Sec.~\ref{sec:conclusions}. 

\section{The $e^+e^-\to\chi\widetilde G$ process}
\label{sec:cross}
\begin{figure}[p]
\vspace{3in}
\includegraphics{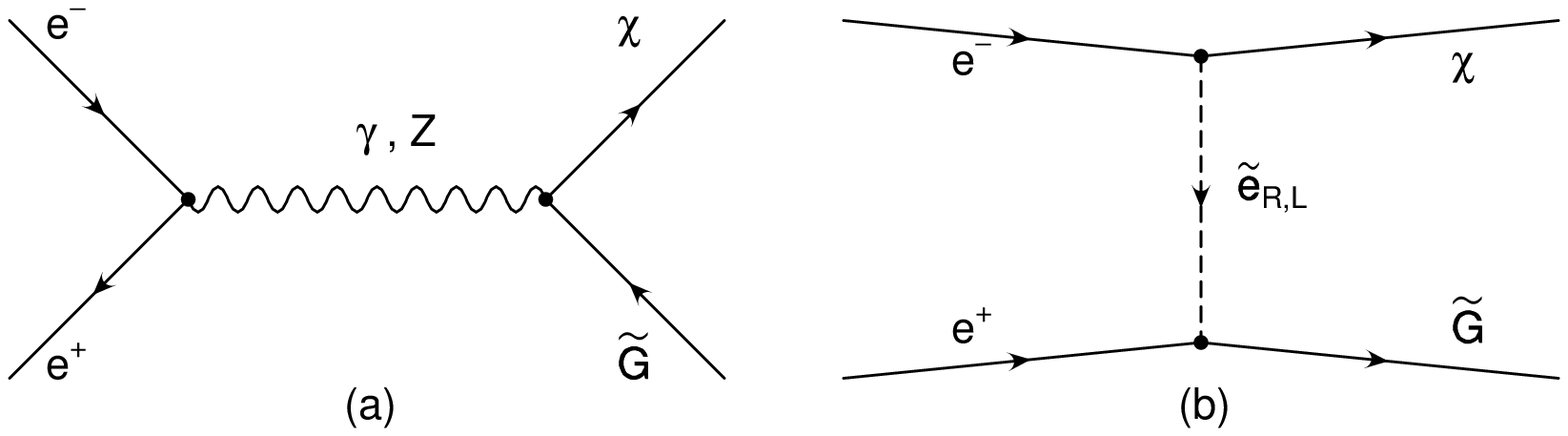}
\caption{Feynman diagrams for neutralino-gravitino production at LEP:
(a) $s$-channel $\gamma$ and $Z$ exchange, and (b) $t$-channel selectron
exchange ($\tilde e_{R,L}$). Additional $u$-channel diagram not shown.}
\label{fig:N1G}
\end{figure}

The Feynman diagrams for neutralino-gravitino associated production at LEP
are shown in Fig.~\ref{fig:N1G}, and include $s$-channel $\gamma$ and $Z$
exchange, and $t$- and $u$-channel selectron ($\tilde e_{R,L}$) exchange.
At the $Z$ peak one expects the $s$-channel $Z$ exchange diagram to dominate,
and one may simply calculate the amplitude for $Z\to\chi\widetilde G$ decay \cite{DicusZ}. This result is accurate as long as the neutralino has a non-negligible zino component. However, for photino-like neutralinos the other diagrams become important. This is also the case for any neutralino composition for center-of-mass energies away from the $Z$ peak ($\sqrt{s}>M_Z$). To deal with all cases at once, we perform the complete calculation of all diagrams contributing to $e^+e^-\to\chi\widetilde G$. We first present the general form of the differential cross section and later specialize the result for the particular case of a photino-like neutralino in order to expound on certain theoretical issues and generalizations of our results.

In calculating interactions of gravitinos with matter one can proceed in one
of two ways. One may calculate with the full couplings in the supergravity Lagrangian, in which case the vertices of interest are given by 
\cite{FayetEarly,MoroiGherghetta}
\begin{equation}
e\ \tilde e_{R,L}\ \widetilde G_\mu \propto
{\gamma_\nu\gamma_\mu\over\sqrt{2}M}\ P_{R,L}\ p^\nu_{\tilde e}\quad;
\qquad\qquad
\gamma_\sigma\ \tilde\gamma\ \widetilde G_\mu \propto 
{1\over M}\ p^\rho_\gamma\ [\gamma_\rho,\gamma_\sigma]\ \gamma_\mu\ ,
\label{eq:full}
\end{equation}
where the (spin-$1\over2$ goldstino component of the) gravitino field is 
$\widetilde G_\mu\propto \partial_\mu\psi/m_{\tilde G}$, and $M=2.4\times10^{18}\,{\rm GeV}$ is the appropriately scaled Planck mass. 
Alternatively one may calculate using a set of much-simplified effective goldstino couplings \cite{Fayet}
\begin{equation}
e\ \tilde e_{R,L}\ \psi \propto{m^2_{\tilde e}-m^2_e\over\sqrt{6} Mm_{\tilde G}}\ P_{R,L}
\quad;
\qquad\qquad
\gamma_\sigma\ \tilde\gamma\ \psi \propto 
{m_{\tilde\gamma}\over\sqrt{6}M m_{\tilde G}}\ [\gamma_\rho,\gamma_\sigma]\ 
p^\rho_\gamma\ .
\label{eq:effective}
\end{equation}
The full and effective couplings give the same results for the cross sections of processes where the typical bad high-energy behavior of the gravitational amplitudes is cancelled completely by the diagrams involving only gravitinos and regular supersymmetric particles. This is the case for the $e^+e^-\to\chi\widetilde G$ process in hand, where we have verified (in the pure photino limit) that both ways of doing the calculation give identical results. For a derivation and explanation of the meaning of the effective couplings see Ref.~\cite{MoroiGherghetta}. The simplification of using the effective couplings is not beneficial in other processes, such as gravitino pair-production, where diagrams including graviton exchanges must also be included to cancel the bad high-energy behavior of the amplitudes \cite{Fayet,MoroiGherghetta}. Also, it is not clear whether this simplification may be used in the case of broken gauge symmetries, and therefore we have used the full couplings in the case of neutralino compositions other than pure photino, where the $s$-channel $Z$-exchange amplitude must be taken into account. 

\subsection{General case}
\label{sec:general}
The differential cross section for a general neutralino composition is given by
\begin{equation}
{d\sigma\over d\cos\theta}={(s-m_\chi^2)\over 32\pi s^2}\,
{F(s,t,u)\over 6 (Mm_{\tilde G})^2}\ ,
\label{eq:sigma}
\end{equation}
where as usual we define
\begin{eqnarray}
t&=&-{\textstyle{1\over2}}(s-m_\chi^2)\,(1-\cos\theta)\ ,\label{eq:t}\\
u&=&-{\textstyle{1\over2}}(s-m_\chi^2)\,(1+\cos\theta)\ .\label{eq:u}
\end{eqnarray}
The function $F(s,t,u)$ receives contributions from each amplitude squared and
various interference terms (some of which vanish). In an obvious notation, we find
\begin{equation}
F=F_{\gamma\gamma}+F_{tt}+F_{uu}+F_{ZZ}
+F_{\gamma t}+F_{\gamma u}+F_{Zt}+F_{Zu}+F_{\gamma Z}\ ,
\label{eq:F}
\end{equation}
where
\begin{eqnarray}
F_{\gamma\gamma}&=&(N'_{11}e)^2\,{2s(s-m^2_\chi)(t^2+u^2)\over s^2}
\label{eq:Fgg}\\
F_{tt}&=&(X_R)^2\,{t^2\,(m^2_\chi-t)(-t)\over(t-m^2_{\tilde e_R})^2}
+(X_L)^2\,{t^2\,(m^2_\chi-t)(-t)\over(t-m^2_{\tilde e_L})^2}
\label{eq:Ftt}\\
F_{uu}&=&(X_R)^2\,{u^2\,(m^2_\chi-u)(-u)\over(u-m^2_{\tilde e_R})^2}
+(X_L)^2\,{u^2\,(m^2_\chi-u)(-u)\over(u-m^2_{\tilde e_L})^2}
\label{eq:Fuu}\\
F_{ZZ}&=&\left(N'_{12}\,{g\over\cos\theta_W}\right)^2\,{(c^2_R+c^2_L)\over2}\,
{2s(s-m^2_\chi)(t^2+u^2)\over (s-M^2_Z)^2+(\Gamma_Z M_Z)^2}
\label{eq:FZZ}\\
F_{\gamma t}&=&(N'_{11}\,eX_R){(-t)(2st^2)\over s(t-m^2_{\tilde e_R})}
-(N'_{11}eX_L){(-t)(2st^2)\over s(t-m^2_{\tilde e_L})}
\label{eq:Fgt}\\
F_{\gamma u}&=&(N'_{11}\,eX_R){(-u)(2su^2)\over s(u-m^2_{\tilde e_R})}
-(N'_{11}eX_L){(-u)(2su^2)\over s(u-m^2_{\tilde e_L})}
\label{eq:Fgu}\\
F_{Zt}&=&-\left(N'_{12}\,c_R X_R\,{g\over\cos\theta_W}\right)
{(-t)(2st^2)(s-M^2_Z)\over[(s-M^2_Z)^2+(\Gamma_Z M_Z)^2](t-m^2_{\tilde e_R})}
\nonumber\\
&&+\left(N'_{12}\,c_L X_L\,{g\over\cos\theta_W}\right)
{(-t)(2st^2)(s-M^2_Z)\over[(s-M^2_Z)^2+(\Gamma_Z M_Z)^2](t-m^2_{\tilde e_L})}
\label{eq:FZt}\\
F_{Zu}&=&-\left(N'_{12}\,c_R X_R\,{g\over\cos\theta_W}\right)
{(-u)(2su^2)(s-M^2_Z)\over[(s-M^2_Z)^2+(\Gamma_Z M_Z)^2](u-m^2_{\tilde e_R})}
\nonumber\\
&&+\left(N'_{12}\,c_L X_L\,{g\over\cos\theta_W}\right)
{(-u)(2su^2)(s-M^2_Z)\over[(s-M^2_Z)^2+(\Gamma_Z M_Z)^2](u-m^2_{\tilde e_L})}
\label{eq:FZu}\\
F_{\gamma Z}&=&-2\left(N'_{11}N'_{12}\,e\,{g\over\cos\theta_W}\right)\,{(c_R+c_L)\over2}\,{2s(s-m^2_\chi)(t^2+u^2)(s-M^2_Z)\over s[(s-M^2_Z)^2+(\Gamma_Z M_Z)^2]}
\label{eq:FgZ}
\end{eqnarray}
In these expressions we have used the following $e$--$\tilde e_{R,L}$--$\chi$ 
($X_{R,L}$) and $e$--$e$--$Z$ ($c_{R,L}$) couplings \cite{HK}
\begin{eqnarray}
X_R&=&N'_{11}\,e-N'_{12}\, {g\sin^2\theta_W\over\cos\theta_W}\ ,\label{eq:XR}\\
X_L&=&-N'_{11}\,e-N'_{12}\,{g\over\cos\theta_W}\,
({\textstyle{1\over2}}-\sin^2\theta_W)\ ,\label{eq:XL}\\
c_R&=&\sin^2\theta_W\ , \label{eq:cR}\\
c_L&=&-{\textstyle{1\over2}}+\sin^2\theta_W\ , \label{eq:cL}
\end{eqnarray}
where $N'_{11}$ and $N'_{12}$ denote the photino and zino components of the
neutralino respectively. Indeed, the lightest neutralino may be written in two equivalent ways \cite{HK}:
\begin{eqnarray}
\chi&=&N'_{11}\widetilde\gamma+N'_{12}\widetilde Z+N_{13}\widetilde H^0_1+N_{14}\widetilde H^0_2\\
\chi&=&N_{11}\widetilde B+N_{12}\widetilde W_3+N_{13}\widetilde H^0_1+N_{14}\widetilde H^0_2
\end{eqnarray}
related by
\begin{equation}
N'_{11}=N_{11}\cos\theta_W+N_{12}\sin\theta_W\ ,\qquad 
N'_{12}=-N_{11}\sin\theta_W+N_{12}\cos\theta_W
\end{equation}

\subsection{Pure photino case}
\label{sec:photino}
This special case is useful in order to expose various subtleties in the calculation that become less apparent (although they are still present) in the case of a general neutralino composition. The $e^+e^-\to\tilde\gamma\widetilde G$ case is also important because the result can be readily taken over to the case of gluino-gravitino production in quark-antiquark annihilation at hadron colliders ($q\bar q\to\tilde g\widetilde G$).

In this special case the couplings of the neutralino (photino: $N'_{11}=1$, $N'_{12}=0$) to matter are very simple: $X_R=e=-X_L$, $F_{ZZ}=F_{Zt}=F_{Zu}=F_{\gamma Z}=0$, and
\begin{equation}
F\to F^{\tilde\gamma}=e^2\,(F^{\tilde\gamma}_\gamma+F^{\tilde\gamma}_R
+F^{\tilde\gamma}_L) \ ,
\label{eq:Fphotino}
\end{equation}
where
\begin{eqnarray}
F^{\tilde\gamma}_\gamma&=&{2s(s-m^2_\chi)(t^2+u^2)\over s^2}
\label{eq:Fphotinog}\\
F^{\tilde\gamma}_R&=&{t^2\,(m^2_\chi-t)(-t)\over(t-m^2_{\tilde e_R})^2}
+{u^2\,(m^2_\chi-u)(-u)\over(u-m^2_{\tilde e_R})^2}
+{t^2(-2st)\over s(t-m^2_{\tilde e_R})}
+{u^2(-2su)\over s(u-m^2_{\tilde e_R})}
\label{eq:FphotinoR}\\
F^{\tilde\gamma}_L&=&{t^2\,(m^2_\chi-t)(-t)\over(t-m^2_{\tilde e_L})^2}
+{u^2\,(m^2_\chi-u)(-u)\over(u-m^2_{\tilde e_L})^2}
+{t^2(-2st)\over s(t-m^2_{\tilde e_L})}
+{u^2(-2su)\over s(u-m^2_{\tilde e_L})}
\label{eq:FphotinoL}
\end{eqnarray}

With this relatively simple expression we can verify certain expected behaviors
of the cross section. First, in the limit of unbroken supersymmetry:
$m_\chi\to m_\gamma=0$, $m_{\tilde e_{L,R}}\to m_e\approx0$, $s+t+u=m^2_\chi\to0$, one can readily verify from the above equations that $F^{\tilde\gamma}\to0$. The vanishing of the cross section in this limit is expected as the spin-$1\over2$ component of the gravitino (the goldstino) becomes an unphysical particle when supersymmetry is unbroken, as it is no longer absorbed by the gravitino to become massive. 

A related manifestation of this phenomenon can be exposed by studying the
threshold behavior of the cross section. The spin-$1\over2$ (goldstino) component of the gravitino is essentially obtained by taking the derivative of the full gravitino field, thus making the goldstino couplings be proportional
to the goldstino momentum ($k^\mu$). At threshold $k^\mu\to0$ and there is
an additional suppression of the cross section besides the kinematical one.
Threshold corresponds to the limit $s\to m^2_\chi$ and therefore from Eqs.~(\ref{eq:t},\ref{eq:u}) $t,u$ go to zero as $(s-m^2_\chi)$. In the above expression for $F^{\tilde\gamma}$, one can see that near threshold each term is proportional to $(s-m^2_\chi)^3$, which combined with the $(s-m^2_\chi)$ term
from the phase space integration [see Eq.~(\ref{eq:sigma})] yields a cross section proportional to $\beta^8$ with $\beta=\sqrt{1-m^2_\chi/s}$. 

The above results were originally obtained by Fayet \cite{Fayet} based on a
calculation of the cross section using the effective couplings (Eq.~(\ref{eq:effective})), and have been obtained here for the first time using the full couplings (Eq.~(\ref{eq:full})). Such equivalent
expression for the cross section makes more evident some further properties
of the results, and we thus give it explicitly here too. The expression
for $F^{\tilde\gamma}$ using the effective couplings becomes
\begin{eqnarray}
F^{\tilde\gamma}_{\rm eff}&=&
m^4_\chi\, {2s(s-m^2)+4uts/m^2_\chi\over s^2}\nonumber\\
&+&m^4_{\tilde e_R}\left[{(m^2_\chi-t)(-t)\over(t-m^2_{\tilde e_R})^2}
+{(m^2_\chi-u)(-u)\over(u-m^2_{\tilde e_R})^2}\right]
+m^2_\chi m^2_{\tilde e_R}\left[{(-2st)\over s(t-m^2_{\tilde e_R})}
+{(-2su)\over s(u-m^2_{\tilde e_R})}\right]\nonumber\\
&+&m^4_{\tilde e_L}\left[{(m^2_\chi-t)(-t)\over(t-m^2_{\tilde e_L})^2}
+{(m^2_\chi-u)(-u)\over(u-m^2_{\tilde e_L})^2}\right]
+m^2_\chi m^2_{\tilde e_L}\left[{(-2st)\over s(t-m^2_{\tilde e_L})}
+{(-2su)\over s(u-m^2_{\tilde e_L})}\right]\nonumber\\
\label{eq:Feff}
\end{eqnarray}
Despite the seemingly different appearances of $F^{\tilde\gamma}_{\rm eff}$
and $F^{\tilde\gamma}$, it can be verified (at least numerically) that they
give identical results. Using $F^{\tilde\gamma}_{\rm eff}$ it is immediately
apparent that the cross section vanishes in the unbroken supersymmetry limit
({\em i.e.}, $m_\chi,m_{\tilde e_{R,L}}\to0$), as it should. (Note also that
for a massless photino (a case of interest in the early literature) the
$s$-channel diagram does not contribute.) The $\beta^8$ threshold behavior is not so apparent this time. One can first note that near threshold $F^{\tilde\gamma}_{\rm eff}$ becomes independent of $m_{\tilde e_{R,L}}$ and depends only on $m^2_\chi$. A little algebra then shows that indeed, near
threshold, $F^{\tilde\gamma}_{\rm eff}\propto(s-m^2_\chi)^3$, and thus the
same $\beta^8$ threshold behavior results, although this time as a result
of a cancellation among all of the contributing amplitudes.

The $F^{\tilde\gamma}_{\rm eff}$ form is also useful in exhibiting the dependence of the cross section on the selectron masses. As is evident from
Eq.~(\ref{eq:Feff}), the cross section increases with increasing selectron
masses, eventually saturating for very large values of $m_{\tilde e}$. Thus, the decoupling theorem still holds ({\em i.e.}, large values of the sparticle masses have no effect), although its specific implementation here is rather peculiar.

Before moving on to numerical evaluations of the cross sections, let us note
that the above expressions for the photino cross section (using either the
full or effective couplings) can be adapted very easily to describe gluino-gravitino production in quark-antiquark collisions at the Tevatron or LHC ($q\bar q\to\tilde g\widetilde G$). In this case the process is mediated
by $s$-channel gluon exchange and $t$-channel $\tilde q_{L,R}$ exchange.
One needs to replace the $e$--$\tilde e_{R,L}$--$\chi$ ($X_{R,L}$) couplings in Eqs.~(\ref{eq:XR},\ref{eq:XL}) by those appropriate for 
$q$--$\tilde q_{R,L}$--$\chi$, one needs to replace the $e$--$e$--$\gamma$
coupling ($e^2$ in Eq.~(\ref{eq:Fphotino})) by the strong coupling ($g^2_s$),
and one needs to insert the appropriate color factor. Of course the integration
over parton distribution functions also needs to be implemented. (A realistic
calculation would also include the gluon-fusion channel, which becomes quite
relevant at LHC energies.)

\section{Experimental constraints}
\label{sec:limits}
\subsection{LEP~1}
The single-photon signal ($\gamma+{\rm E_{miss}}$) has been searched for
at LEP~1 by various LEP Collaborations \cite{LEP1}. We estimate an
upper bound of 0.1~pb on this cross section. This estimate is an amalgamation of individual experimental limits with partial LEP~1 luminosities ($\sim100\,{\rm pb}^{-1}$) and angular acceptance restrictions ($|\cos\theta_\gamma|<0.7$). Note that the single-photon background at the $Z$ peak (mostly from $e^+e^-\nu\bar\nu\gamma$) is quite significant, as otherwise one would naively expect upper bounds of order $3/{\cal L}< 0.03\,{\rm pb}$.

\begin{figure}[p]
\vspace{6in}
\includegraphics{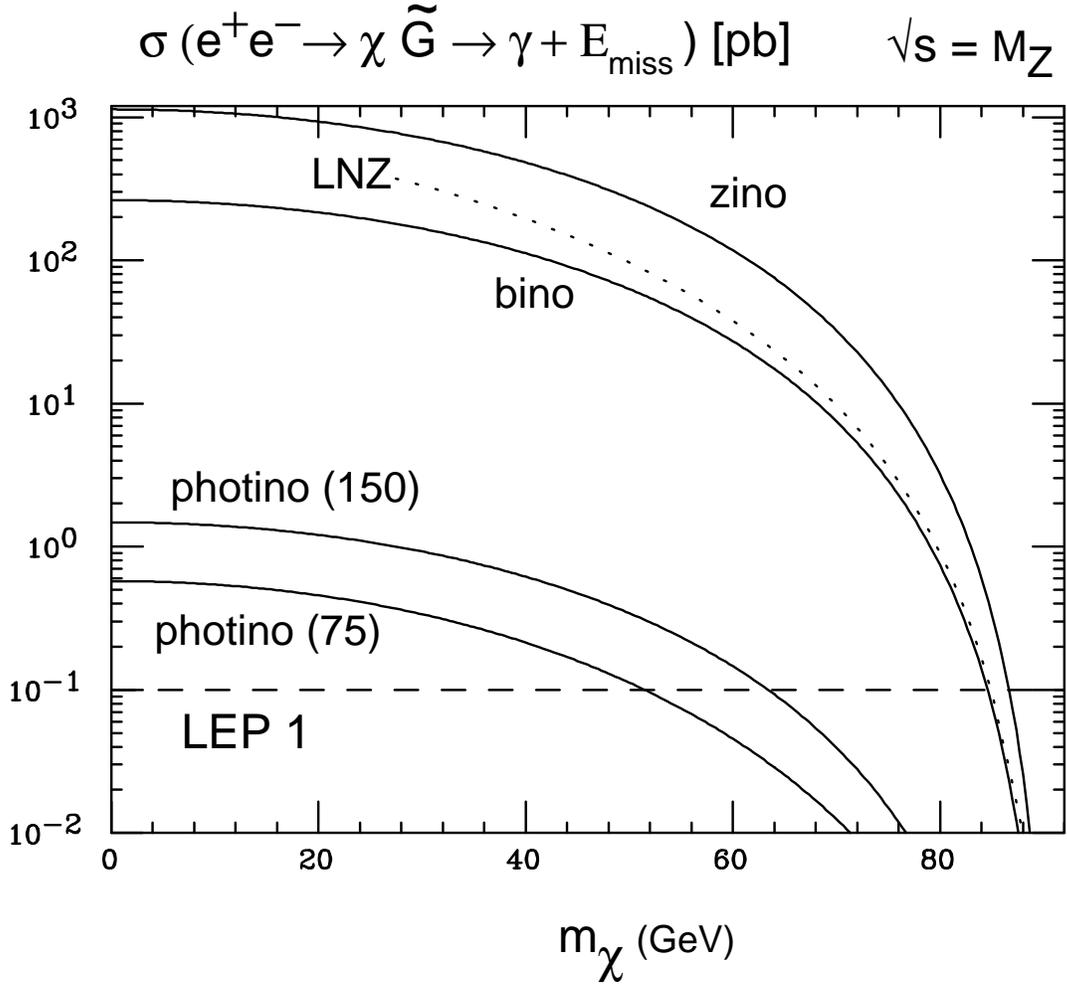}
\caption{Single-photon cross sections (in pb) from neutralino-gravitino
production at LEP~1 versus the neutralino mass ($m_\chi$) for
$m_{\widetilde G}=10^{-5}\,{\rm eV}$ and various neutralino compositions.
The `photino' curves depend on the selectron mass (75,150).
The cross sections scale like $\sigma\propto m^{-2}_{\widetilde G}$. The
dashed line represents the estimated LEP~1 upper limit. Also shown is the
result for a one-parameter no-scale supergravity model (`LNZ').}
\label{fig:sigmaMz}
\end{figure}

A numerical evaluation of the single-photon cross section at LEP~1 versus the neutralino mass for $m_{\widetilde G}=10^{-5}\,{\rm eV}$ is shown in Fig.~\ref{fig:sigmaMz}, for different choices of neutralino composition (`zino': $N'_{12}\approx1$; `bino': $N_{11}=1$, and `photino': $N'_{11}=1$),
and where we have assumed the typical result 
$B(\chi\to\gamma\widetilde G)=1$.\footnote{Note that this implies a  non-vanishing (possibly small) photino component of the neutralino, as would
be required in the `zino' case discussed above.} In the photino case the $Z$-exchange amplitude is absent ($N'_{11}=1\Rightarrow N'_{12}=0$) and one must also specify the selectron masses which mediate the
$t$- and $u$-channel diagrams; we have taken the representative values 
$m_{\tilde e_R}=m_{\tilde e_L}=75,150\,{\rm GeV}$. Increasing the selectron
masses further leads to only a small increase in the cross section, {\em e.g.},
at $m_\chi=0$ one finds  $\sigma^{\rm M_Z}_\gamma=1.48,2.09,2.36\,{\rm pb}$ for 
$m_{\tilde e}=150,300,1000\,{\rm GeV}$, signalling the reaching of the
decoupling limit for large selectron masses discussed in Sec.~\ref{sec:photino}.

In Fig.~\ref{fig:sigmaMz} we also show (dotted line `LNZ') the results for a well-motivated one-parameter no-scale supergravity model \cite{Gravitino,One}, which realizes the light gravitino scenario that we study here. In this model
the neutralino is mostly gaugino, but has a small higgsino component at low values of $m_\chi$, which disappears with increasing neutralino masses;
the neutralino approaches a pure bino at high neutralino masses. The selectron
masses also vary (increase) continously with the neutralino mass and are not
degenerate ({\em i.e.}, $m_{\tilde e_L}\sim 1.5\, m_{\tilde e_R}\sim2 m_\chi$).

This figure makes apparent the constraint on the gravitino mass that arises
from LEP~1 searches: in some regions of parameter space one must require
$m_{\tilde G}\gsim10^{-3}\,{\rm eV}$ if $m_\chi<M_Z$.\footnote{For reference, in gauge-mediated models of low-energy supersymmetry, such gravitino masses correspond to $\Lambda_{\rm SUSY}\sim 3\,{\rm TeV}$.} To make this result
more evident, in Fig.~\ref{fig:limits} we display the lower bound on the gravitino mass versus the neutralino mass, that results from the imposition
of our estimated upper bound $\sigma^{\rm M_Z}_\gamma<0.1\,{\rm pb}$. 
(The curves that extend beyond $m_\chi=M_Z$ result from constraints from
LEP~2 data, and are discussed below.) Note the dependence on the selectron
mass in the pure photino case.

\subsection{LEP161}
\label{sec:LEP161}
Recent runs of LEP at higher center-of-mass energies have so far yielded no
excess of single photons over Standard Model expectations. The latest searches
at $\sqrt{s}=161\,{\rm GeV}$ have produced upper limits on the single-photon
cross section $\sigma^{\rm 161}_\gamma\lsim1\,{\rm pb}$ \cite{DELPHI}.
We have evaluated the single-photon cross sections for the neutralino
compositions used in  Fig.~\ref{fig:sigmaMz} at $\sqrt{s}=161\,{\rm GeV}$. This
time all cases depend on the choice of selectron masses. The numerical
results are shown in Fig.~\ref{fig:sigma161}, with the experimental upper
bound denoted by the dashed line. Note that this line extends only for
$m_\chi>M_Z$, as for $m_\chi<M_Z$ the much stronger limits discussed in the previous section apply. Moreover, for $m_{\tilde G}=10^{-5}\,{\rm eV}$ (as used in Fig.~\ref{fig:sigma161}), LEP~1 limits require $m_\chi\gsim M_Z$. As the
figure makes evident, for $m_\chi<M_Z$ the sensitivity to single-photon signals at LEP~2 is not competitive with that at LEP~1. 

As discussed above, the cross sections in Fig.~\ref{fig:sigma161} increase with  increasing selectron masses (saturating at values somewhat larger than the ones shown), and conversely decrease with decreasing selectron masses. The choice of selectron masses also affects the near-threshold behavior of the cross section, with light selectron masses ``delaying" the onset of the $\beta^8$ threshold dependence (see Fig.~\ref{fig:sigma161}). Note also that the photino, bino, and zino cross sections become comparable above the $Z$ pole, when the $Z$-exchange diagram becomes comparable to the other diagrams. In the case of the one-parameter model (`LNZ') a peculiar bump appears. This bump is understood in terms of the selectron masses that vary continously with the neutralino mass: at low values of $m_\chi$ the selectron masses are light and the cross section approaches the light fixed-selectron mass curves (`75'); at larger values of $m_\chi$ the selectron masses are large and the cross section approaches (and exceeds) the heavy fixed-selectron mass curves (`150'). This example brings to light some of the subtle features that might arise in realistic models of low-energy supersymmetry.

In spite of their apparent weakeness, LEP161 limits on the single-photon cross
section are useful in constraining the gravitino mass in a neutralino-mass
range inaccessible at LEP~1. Indeed, decreasing the gravitino mass in Fig.~\ref{fig:sigma161} by a factor of 3 will make the cross sections some
ten-times larger. The resulting lower bounds on the gravitino mass from LEP161
searches are shown in Fig.~\ref{fig:limits}. This figure shows that, as expected, LEP~1 limits dominate for $m_\chi\lsim M_Z$. However, because of the $\beta^8$ threshold behavior at $\sqrt{s}=M_Z$, LEP161 limits `take over' for neutralino masses slightly below $M_Z$, and in the `photino' case, considerably below $M_Z$.

\subsection{Single photons versus diphotons}
It has been made apparent in Fig.~\ref{fig:limits} that for $m_\chi\lsim M_Z$
the gravitino mass is constrained to  $m_{\tilde G}\gg 10^{-5}\,{\rm eV}$.
If this would indeed be an absolute requirement on the gravitino mass ({\em i.e.}, for all values of $m_\chi$) then the cross section for neutralino-gravitino production at LEP~2 would be highly suppressed: Fig.~\ref{fig:sigma161} shows that $\sigma^{\rm 161}_\gamma\lsim1\,{\rm pb}$
for $m_{\tilde G}=10^{-5}\,{\rm eV}$, and $\sigma_\gamma\propto m^{-2}_{\tilde G}$. In other words, if the minimum observable single-photon cross section
at LEP~2 is $\sim0.1\,{\rm pb}$ ({\em i.e.}, for ${\cal L}\sim100\,{\rm pb}^{-1}$), then $m_{\tilde G}\gsim 3\times10^{-5}\,{\rm eV}$ appears to
be the limit of sensitivity of LEP~2.

On the other hand, the process $e^+e^-\to\chi\chi\to\gamma\gamma+{\rm E_{miss}}$ is sensitive to $m_\chi<{1\over2}\sqrt{s}$ and is independent
of the gravitino mass. In light of the single-photon constraints on the
gravitino mass obtained above, the diphoton process may be observable at
LEP~2 ({\em i.e.}, $m_\chi<{1\over2}\sqrt{s}\lsim M_Z$)
only if $m_{\tilde G}\gg10^{-5}\,{\rm eV}$, and therefore single photons
will not be simultaneously observable at LEP~2. Conversely, single photons may be observable at LEP~2 only if $m_\chi>M_Z$, in which case diphotons may not be observed simultaneously (as they require $\sqrt{s}> 2M_Z\approx190\,{\rm GeV}$). This dichotomy between single-photon and diphoton signals at LEP
was first presented in Ref.~\cite{1vs2gamma}.

\subsection{Other limits}
\begin{figure}[p]
\vspace{6in}
\includegraphics{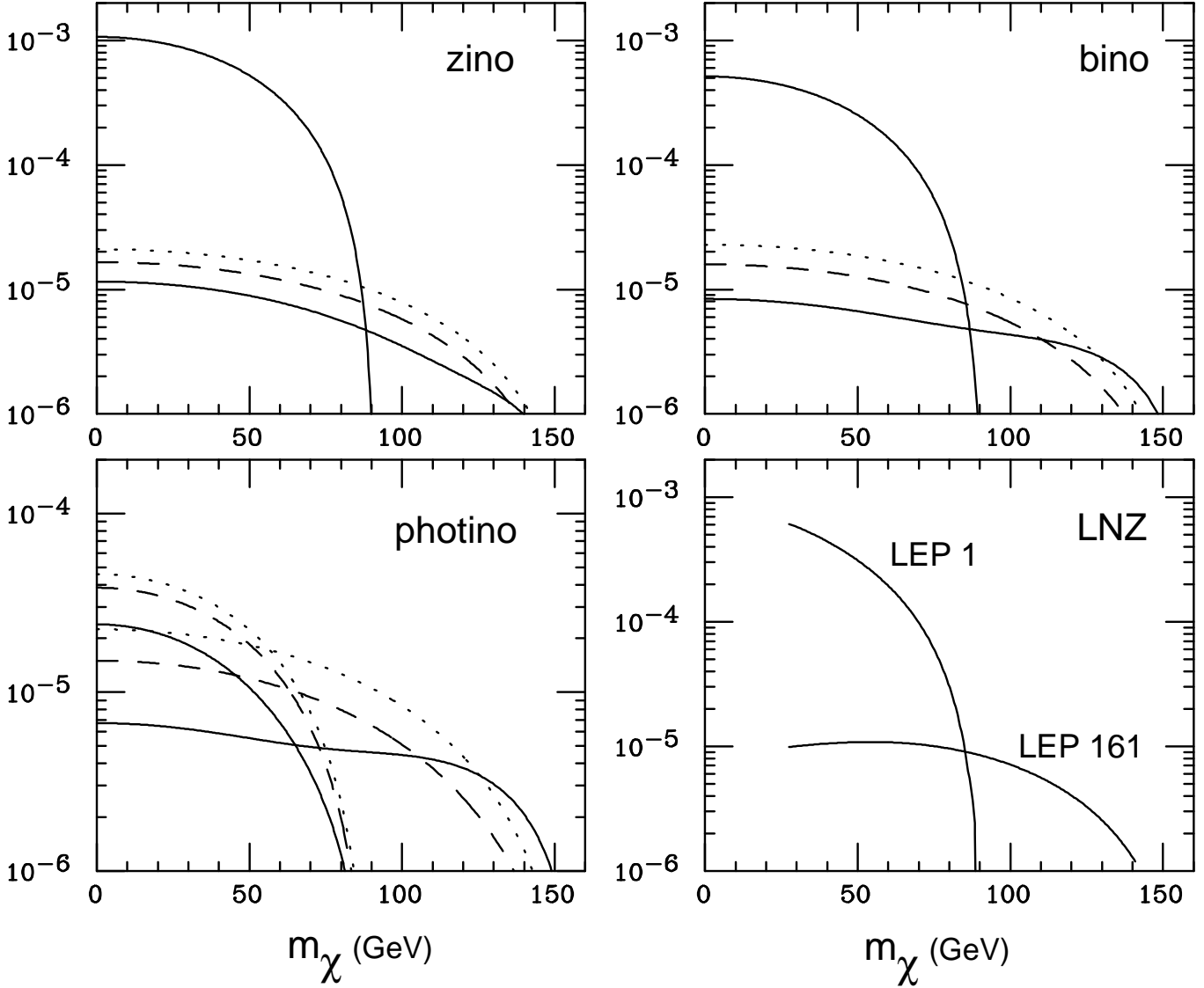}
\vspace{1.5cm}
\caption{Lower bounds on the gravitino mass (in eV) as a function of the
neutralino mass ($m_\chi$) that result from  single-photon searches ($\gamma+{\rm E_{miss}}$) at LEP~1 and LEP161. In the `photino' case at LEP~1
and the `photino', `zino', and `bino' cases at LEP161, the selectron mass 
influences the results. We have chosen $m_{\tilde e}=75,150,300\,{\rm GeV}$,
denoted by solid, dashed, and dotted lines respectively. Also shown are the
bounds in a one-parameter no-scale supergravity model (`LNZ').}
\label{fig:limits}
\end{figure}

\begin{figure}[p]
\vspace{6in}
\includegraphics{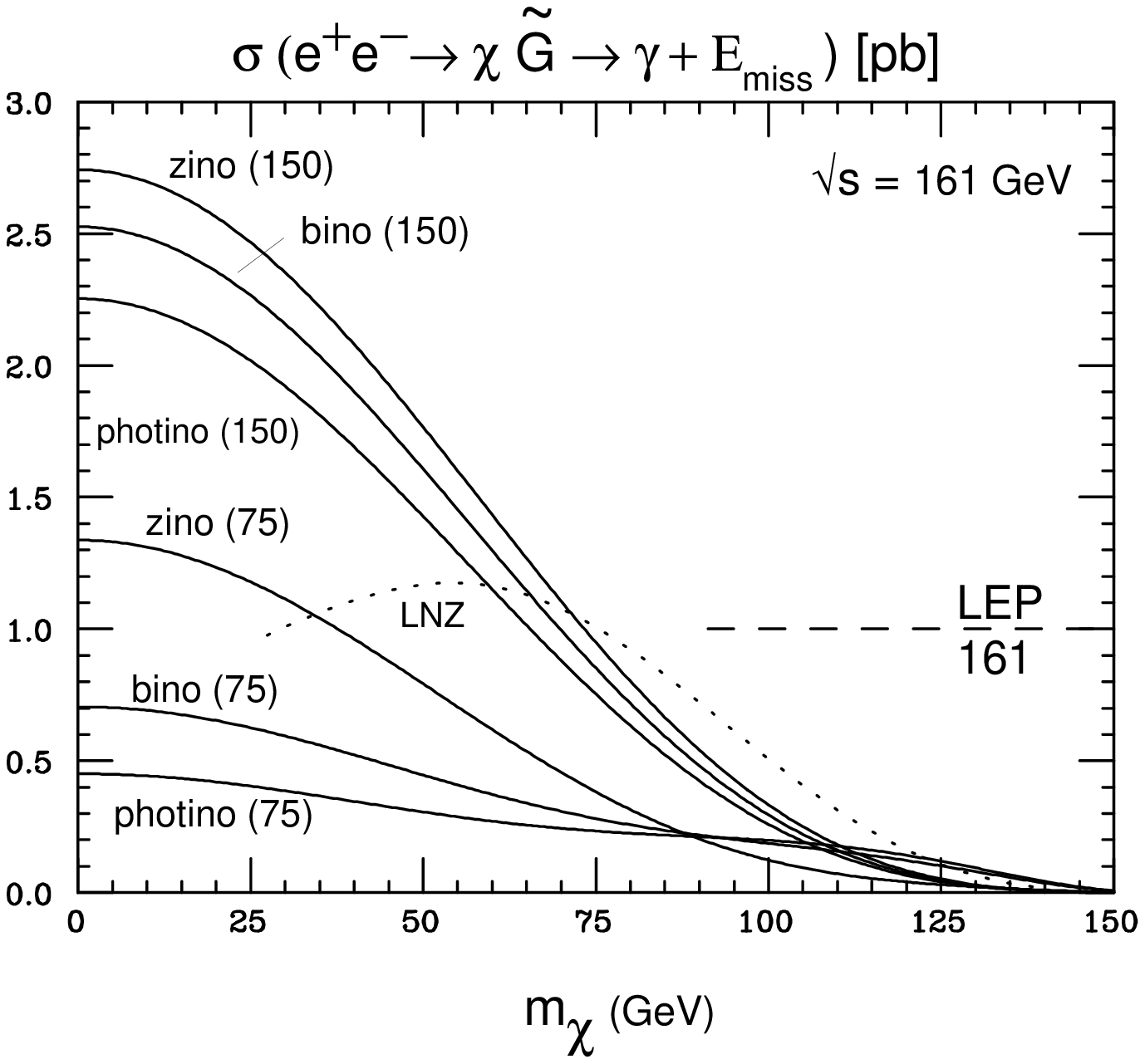}
\caption{Single-photon cross sections (in pb) from neutralino-gravitino
production at LEP~161 versus the neutralino mass ($m_\chi$) for $m_{\widetilde G}=10^{-5}\,{\rm eV}$ and various neutralino compositions.
The solid curves have a fixed value for the selectron mass (75,150), whereas the dotted curve corresponds to a one-parameter no-scale supergravity
model where the selectron masses vary continously with the neutralino mass.
The cross sections scale like $\sigma\propto m^{-2}_{\widetilde G}$. The
preliminary LEP161 upper limit is indicated.}
\label{fig:sigma161}
\end{figure}

The above lower limits on $m_{\tilde G}$ are rather significant, and improve
considerably on previous limits from collider experiments \cite{Fayet,Dicus,DicusZ,LEP1} and astrophysical considerations \cite{astro},
as long as $m_\chi<M_Z$. There has also been a recent reassessment \cite{DicusNandi} of the hadron collider limits obtained via associated gluino-gravitino production ($p\bar p\to \tilde g\widetilde G$), and via indirect gluino pair-production ($p\bar p\to\tilde g\tilde g$) where in addition to the usual supersymmetry QCD diagrams the gravitino is exchanged in the $t$ and $u$ channels. The multijet signature of these processes has been
contrasted with experimental limits from the most recent Tevatron run to show
that if gluino pair production is accessible at the Tevatron ({\em i.e.}, $m_{\tilde g}\lsim200\,{\rm GeV}$) then a lower limit of $m_{\tilde G}>3\times10^{-3}\,{\rm eV}$ results. This limit and our limits in Fig.~\ref{fig:limits} may be compared by relating the gluino and neutralino masses, as occurs in supergravity models with universal gaugino masses at the unification scale: $m_{\tilde g}\sim 3 m_{\chi^\pm}\sim 6 m_\chi$. Therefore, $m_{\tilde g}\lsim200\,{\rm GeV}$ (as required for the bound in Ref.~\cite{DicusNandi} to apply) corresponds to $m_\chi\lsim35\,{\rm GeV}$. Consulting Fig.~\ref{fig:limits}, we see that the Tevatron limits are stronger
in this neutralino mass range. However, the LEP~1\,(161) limit extends
up to $m_\chi\lsim M_Z\,(2M_W)$, which corresponds to $m_{\tilde g}\lsim550\,(960)\,{\rm GeV}$, which is far from the direct reach of the Tevatron.

By considering the further processes $p\bar p\to gS,gP$, where $S$ and $P$ are very light scalar and pseudo-scalar particles associated with the gravitino, the lower bound on the gravitino mass becomes much less dependent on the gluino mass and can be taken to be $m_{\tilde G}>3\times10^{-4}\,{\rm eV}$
\cite{DicusNandi}. This lower bound is comparable with those obtained above
by considering LEP~1 data. However, this result assumes the existence of
additional light and strongly interacting particles ($S$ and $P$), an
assumption that depends on the detailed nature of the mechanism that leads to a very light gravitino. 

\subsection{New channels}
Another set of channels of interest at the Tevatron consist of the associated
production of gravitinos with neutralinos or charginos 
\begin{equation}
p\bar p\to \chi\widetilde G, \chi^\pm\widetilde G\ ,
\label{eq:newchannels}
\end{equation}
which have the advantage over $p\bar p\to\tilde g\widetilde G$ of much less
phase space suppression. The most basic channel is $q\bar q\to\chi\widetilde G$, which leads to a $\gamma+{\rm E_{T,miss}}$ signal. The cross section for this process can be readily obtained from the expressions given in Sec.~\ref{sec:cross} by replacing the initial state electron-positron pairs
by quark-antiquark pairs, the exchanged selectrons by squarks, and by integrating the resulting expression over parton distribution functions.
We have estimated this cross section and find that it may be quite significant:
up to $85,25,15\,{\rm pb}$ for $m_\chi=50,75,100\,{\rm GeV}$ and $m_{\tilde G}=10^{-5}\,{\rm eV}$, in favorable regions of parameter space. In the best case scenario of a Tevatron upper limit of 0.1~pb ({\em i.e.}, 10 events in ${\cal L}=100\,{\rm pb}^{-1}$), one may conclude that $m_{\tilde G}\gsim (3,1.6,1.2)\times10^{-4}\,{\rm eV}$ for $m_\chi=50,75,100\,{\rm GeV}$. Taken at
face value, these limits are quite competitive with those obtained in
Ref.~\cite{DicusNandi}. At the moment there are no single-photon limits available from CDF nor D0. 

To improve the visibility of the signal, one may want to consider the
$q\bar q\to\chi^\pm \widetilde G$ channel which, depending on the chargino
decay channel, may lead to $\ell^\pm+\gamma+{\rm E_{T,miss}}$ or 
$2j+\gamma+{\rm E_{T,miss}}$ signals. The leptonic signal appears particularly
promising. For all these processes there are some important instrumental
backgrounds that need to be overcome. For instance $p\bar p\to W\to e\nu_e$, where the electron is misidentified as a photon ({\em i.e.}, because of limitations in tracking efficiency), leads to a very large ``single-photon" signal $\sigma\cdot B(W\to e\nu_e)\approx 2.4\times10^3\,{\rm pb}$ \cite{Wenu}, which may be reduced significantly by optimizing the tracking efficiency and
making suitable kinematical cuts. The other channels mentioned above face
similar, although perhaps less severe, instrumental backgrounds ({\em e.g.},
$WW\to e+``\gamma"+{\rm E_{T,miss}}$).

\section{The single-photon signal}
\label{sec:signal}
The total cross section for neutralino-gravitino production has been displayed
in Fig.~\ref{fig:sigma161} for a specific center-of-mass energy ($\sqrt{s}=161\,{\rm GeV}$) and for some illustrative choices of parameter
values. The analytic expressions given in Sec.~\ref{sec:general} allow one
to calculate these cross sections for arbitrary values of the parameters.
In this section we would like to explore some characteristics of the actual signal, {\em i.e.}, the energy and angular distributions of the
observable photon and the missing invariant mass distribution in the events.

\subsection{Monte Carlo technique}
Our simulation proceeds in a standard way, making use of a `home-made' Monte
Carlo event generator. We start in the rest frame of the decaying neutralino,
where we generate $\gamma+\widetilde G$ events that are isotropic in this
reference frame. Energy-momentum conservation requires $E'_\gamma=|\vec p{\,'}_\gamma|={1\over2}\,m_\chi$, which leaves  two components of the photon
momentum to be generated at random ({\em i.e.}, $\widehat p'_\gamma$). We then boost the photon momentum back to the laboratory frame using the neutralino 4-momentum $(E_\chi,\vec p_\chi)$, whose components are constrained by the kinematics of the $e^+e^-\to\chi\widetilde G$ process:
\begin{equation}
E_\chi={\sqrt{s}\over2}+{m^2_\chi\over2\sqrt{s}}\,,\qquad
|\vec p_\chi|={\sqrt{s}\over2}-{m^2_\chi\over2\sqrt{s}}\ .
\label{eq:Epchi}
\end{equation}
Here we have two components of the neutralino momentum unconstrained $(\cos\theta_\chi,\phi_\chi)$. For fixed values of these angles we obtain
$E_\gamma$ and $\cos\theta_\gamma$ distributions, which are purely kinematical
effects. The observable distributions are obtained by varying 
$(\cos\theta_\chi,\phi_\chi)$ and weighing these kinematical distributions with the corresponding dynamical ${1\over\sigma}\,{d\sigma\over d\cos\theta_\chi}$ factors calculable from the expressions given in Sec.~\ref{sec:cross}.

\begin{figure}[p]
\vspace{6in}
\includegraphics{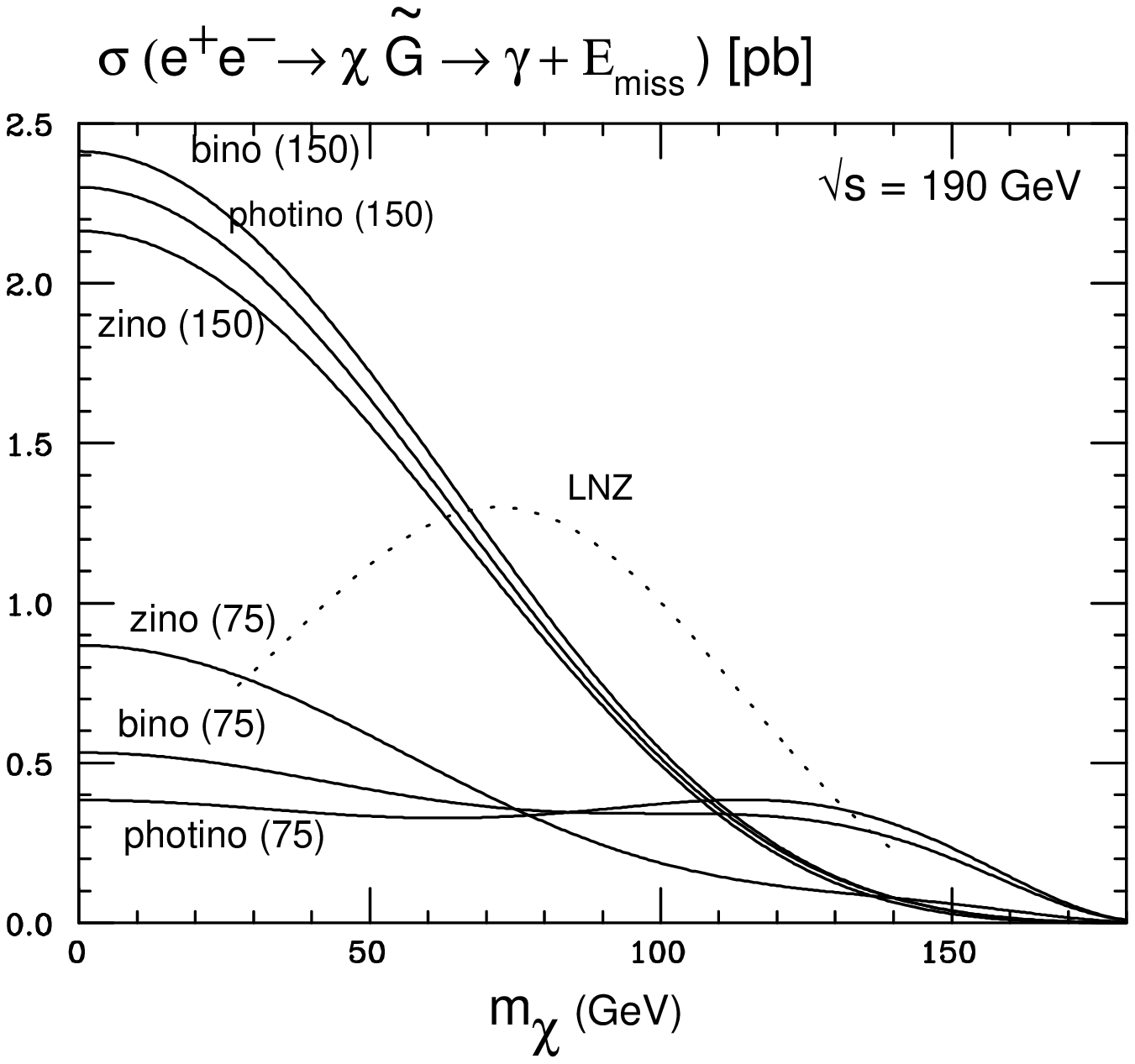}
\caption{Single-photon cross sections (in pb) from neutralino-gravitino
production at LEP~190 versus the neutralino mass ($m_\chi$) for $m_{\widetilde G}=10^{-5}\,{\rm eV}$ and various neutralino compositions.
The solid curves have a fixed value for the selectron mass (75,150), whereas the dotted curve corresponds to a one-parameter no-scale supergravity
model where the selectron masses vary continously with the neutralino mass.
The cross sections scale like $\sigma\propto m^{-2}_{\widetilde G}$.}
\label{fig:sigma190}
\end{figure}

\begin{figure}[p]
\vspace{6in}
\includegraphics{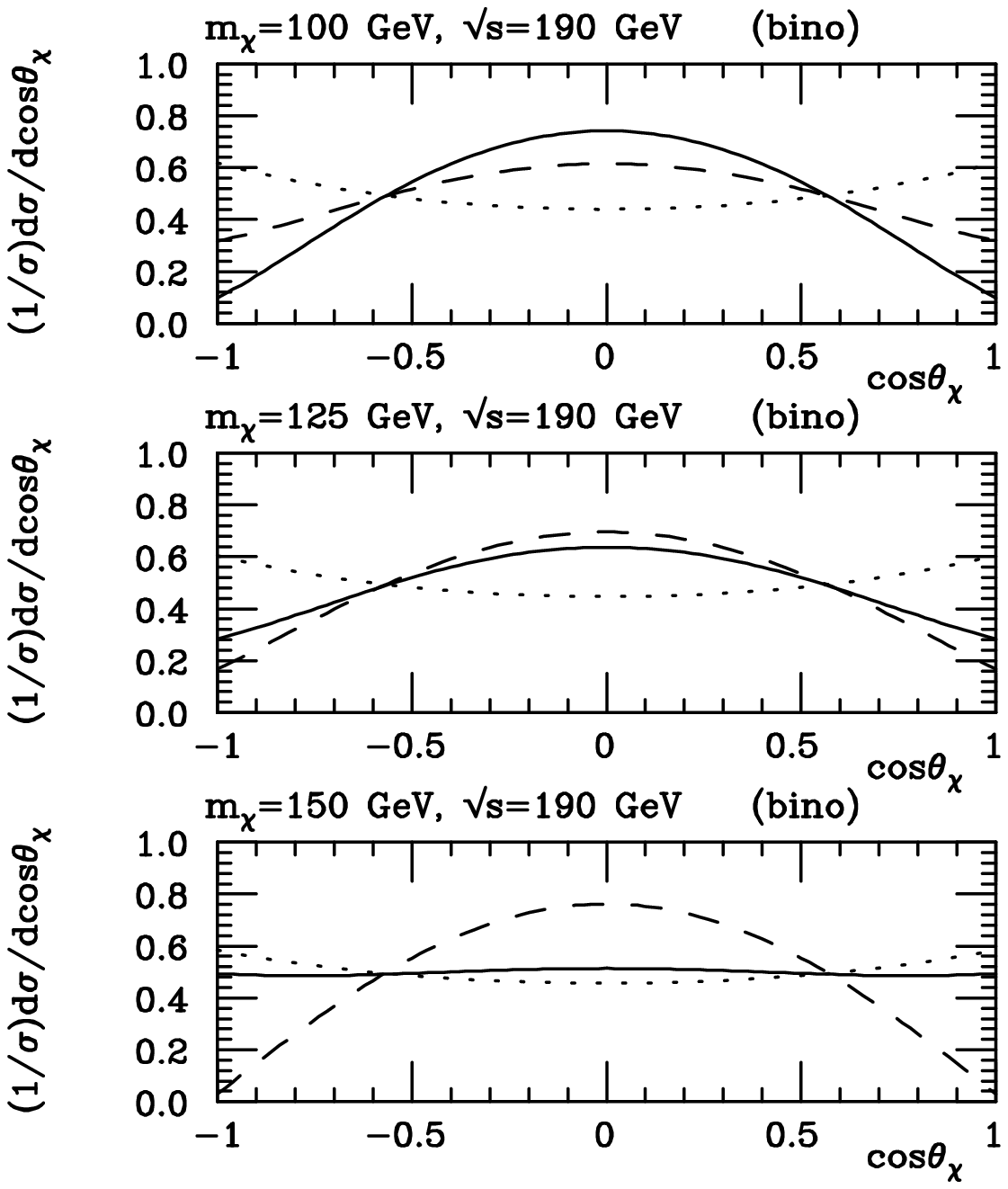}
\caption{Normalized angular distribution of neutralinos of bino composition in neutralino-gravitino production at LEP190 for $m_\chi=100,125,150\,{\rm GeV}$ and $m_{\tilde e}=\rm 75\,(solid),150\,(dashed),300\,(dots)\,{\rm GeV}$.}
\label{fig:Thetachi.bino}
\end{figure}

\begin{figure}[p]
\vspace{6in}
\includegraphics{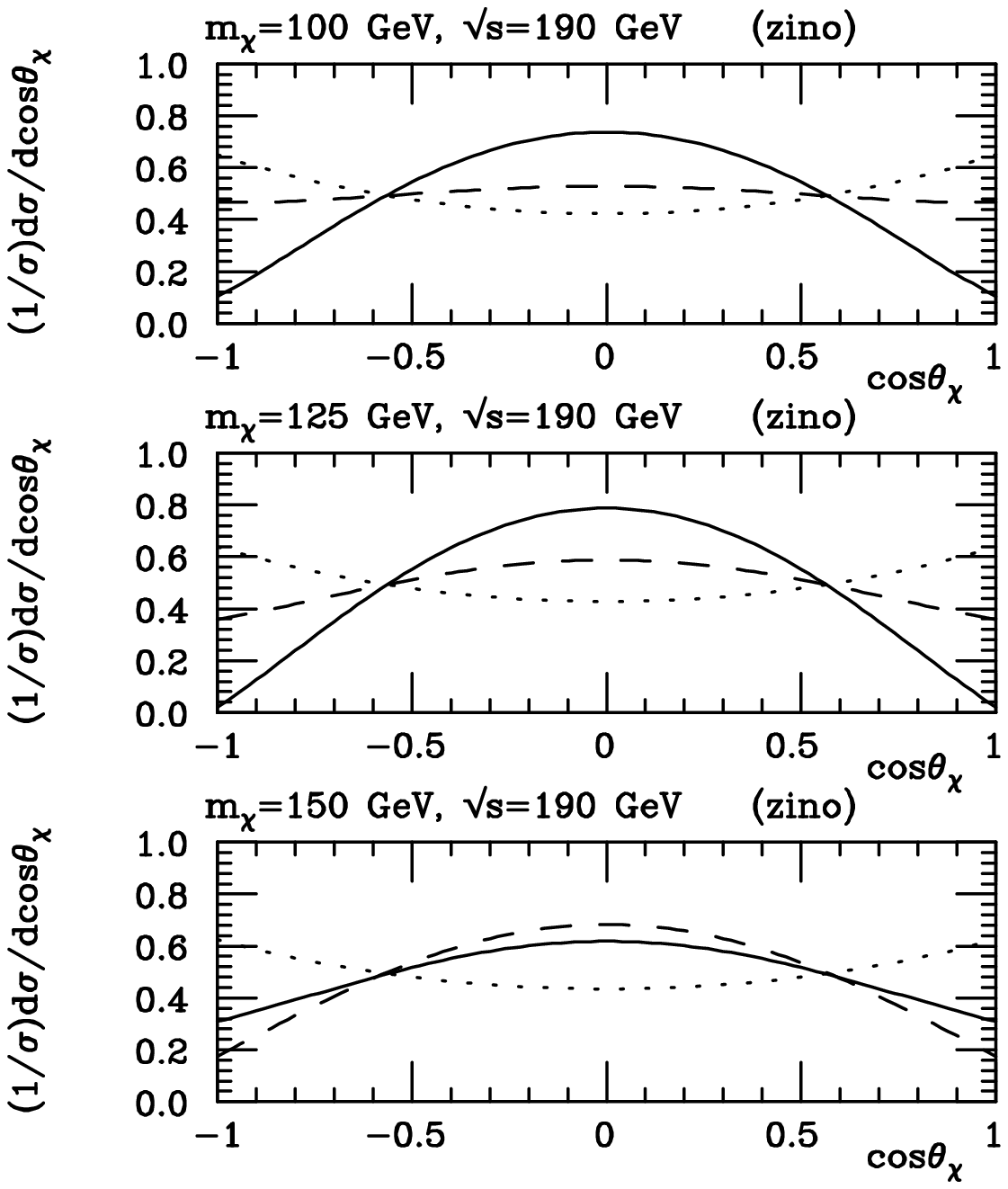}
\caption{Normalized angular distribution of neutralinos of zino composition in neutralino-gravitino production at LEP190 for $m_\chi=100,125,150\,{\rm GeV}$ and $m_{\tilde e}=\rm 75\,(solid),150\,(dashed),300\,(dots)\,{\rm GeV}$.}
\label{fig:Thetachi.zino}
\end{figure}

\begin{figure}[p]
\vspace{6in}
\includegraphics{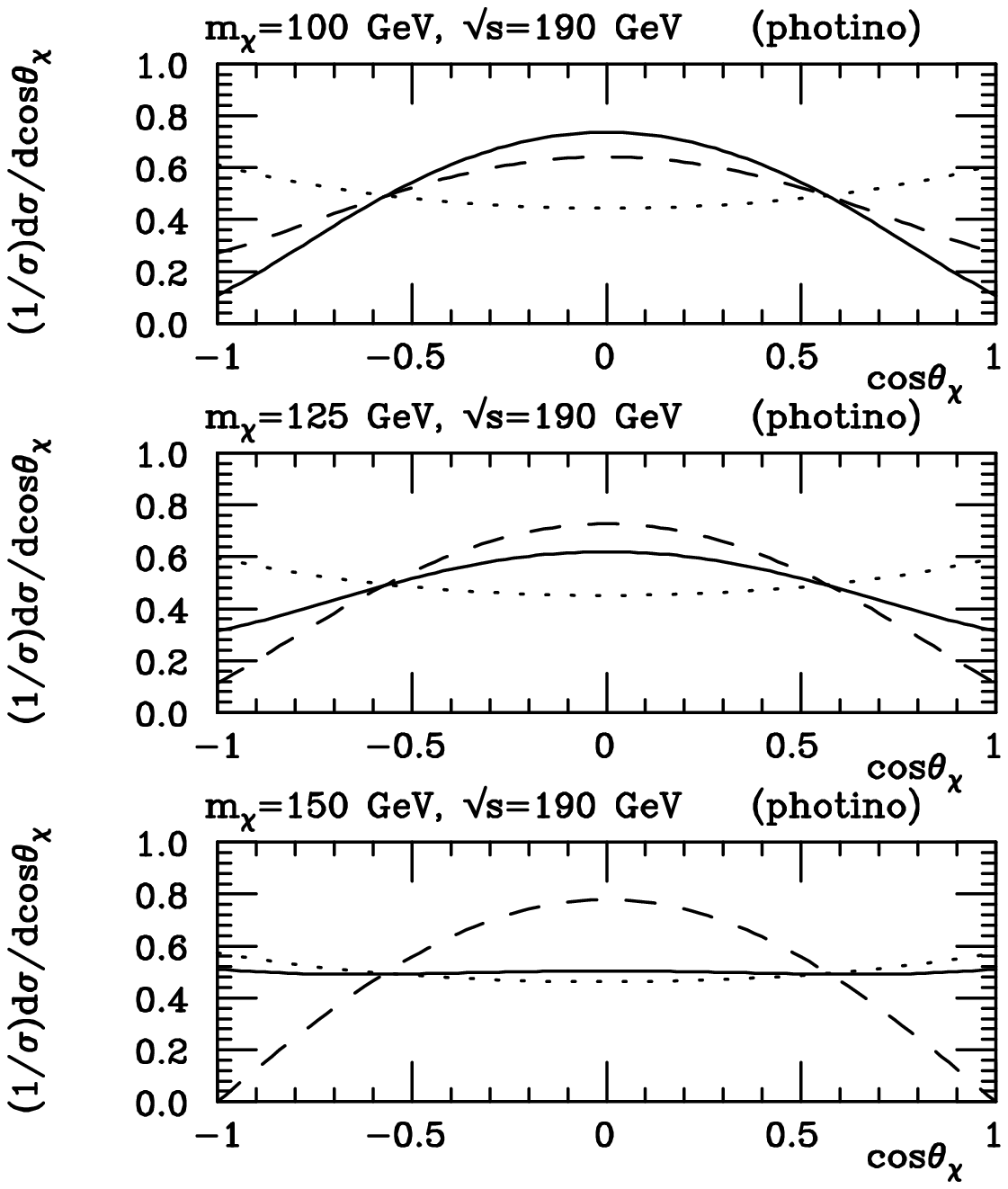}
\caption{Normalized angular distribution of neutralinos of photino composition in neutralino-gravitino production at LEP190 for $m_\chi=100,125,150\,{\rm GeV}$ and $m_{\tilde e}=\rm 75\,(solid),150\,(dashed),300\,(dots)\,{\rm GeV}$.}
\label{fig:Thetachi.photino}
\end{figure}

In what follows we focus on the case of LEP190 ($\sqrt{s}=190\,{\rm GeV}$). First we display in Fig.~\ref{fig:sigma190} the total cross sections for single-photon production at this center-of-mass energy. These should
give an idea of the reach in neutralino masses that may be accessible at LEP190. As we expect neutralino-gravitino production to be allowed only for
$m_\chi>M_Z$ (to avoid the stringent LEP~1 lower limits on $m_{\tilde G}$),
we concentrate on the following three neutralino mass choices: $m_\chi=100,125,150\,{\rm GeV}$. To gain some insight into the final distributions, we start by displaying the normalized neutralino angular distributions ${1\over\sigma}\,{d\sigma\over d\cos\theta_\chi}$ as a function of $\cos\theta_\chi$ for $m_{\tilde e}=75,150,300\,{\rm GeV}$, and bino 
(Fig.~\ref{fig:Thetachi.bino}), zino (Fig.~\ref{fig:Thetachi.zino}), and
photino (Fig.~\ref{fig:Thetachi.photino}) neutralino compositions. The total cross sections for each of the curves can be read off Fig.~\ref{fig:sigma190} and for convenience have been tabulated in Table~\ref{Table1}. As can be
seen from Figs.~\ref{fig:Thetachi.bino}, \ref{fig:Thetachi.zino}, and
\ref{fig:Thetachi.photino}, the angular distribution varies quite a bit
with neutralino mass, although mostly for light selectron masses. Note that
the angular distributions always remain finite, and generally show a preference
for the central region.

\begin{table}[t]
\caption{Total cross sections corresponding to the differential cross sections
shown in Figs. 6,7,8 at LEP190. All masses in GeV, all cross sections in pb.}
\label{Table1}
\begin{center}
\begin{tabular}{lcccc}
Composition&$m_\chi$&$m_{\tilde e}=75$&$m_{\tilde e}=150$&
$m_{\tilde e}=300$\\ \hline
bino&100&0.34&0.54&1.61 \\
&125&0.32&0.19&0.60\\
&150&0.20&0.04&0.11\\ \hline
zino&100&0.19&0.49&1.12\\
&125&0.10&0.17&0.42\\
&150&0.06&0.03&0.08\\ \hline
photino&100&0.37&0.52&1.64\\
&125&0.37&0.18&0.61\\
&150&0.23&0.04&0.12\\
\end{tabular}
\hrule
\end{center}
\end{table}

\subsection{Energy and angular distributions}
\begin{figure}[p]
\vspace{6in}
\includegraphics{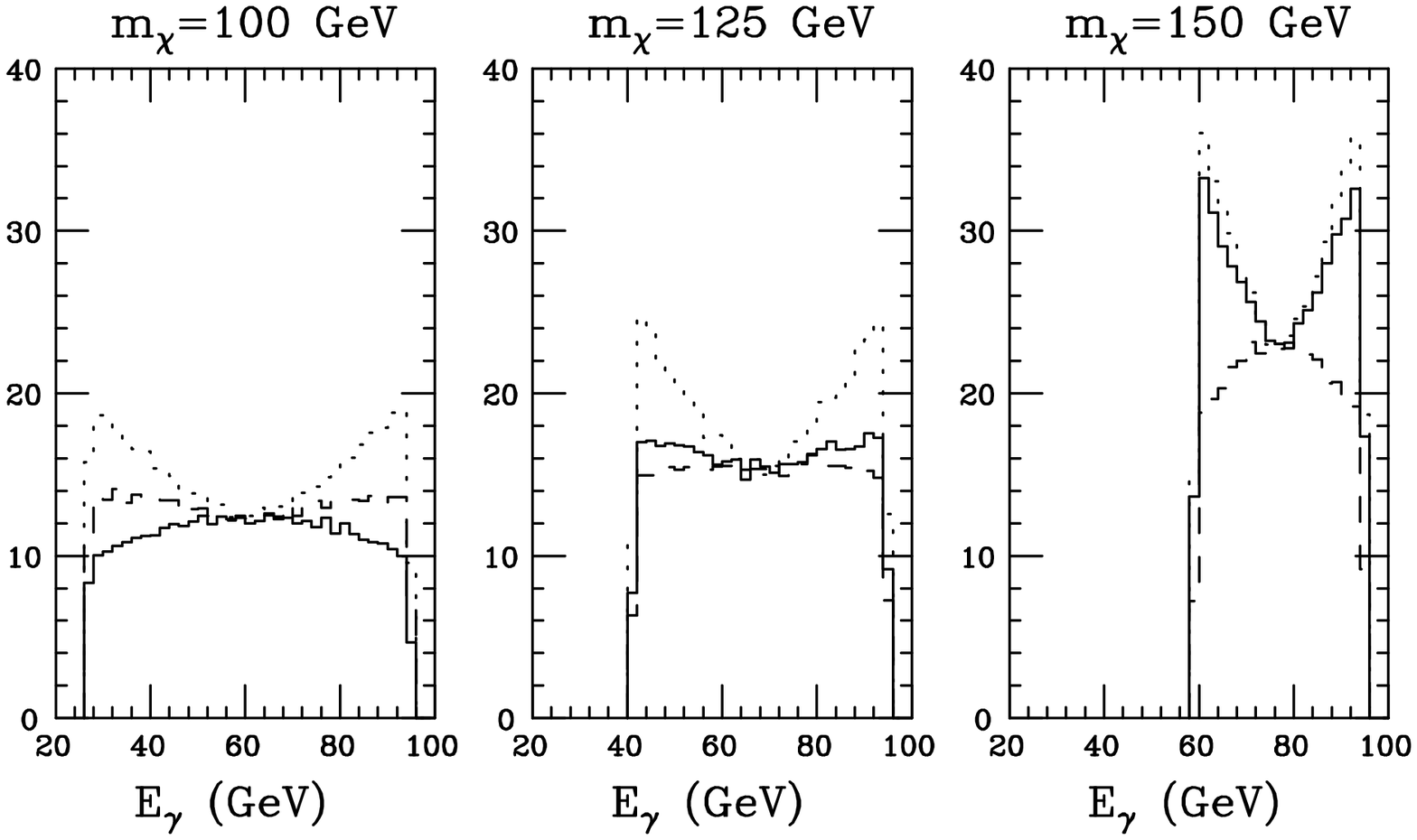}
\caption{Photonic energy distributions in neutralino(bino)-gravitino production at LEP190 for $m_\chi=100,125,150\,{\rm GeV}$ and $m_{\tilde e}=\rm 75$ (solid), 150 (dashed), and 300 (dots) GeV.}
\label{fig:Ebins}
\end{figure}

\begin{figure}[p]
\vspace{6in}
\includegraphics{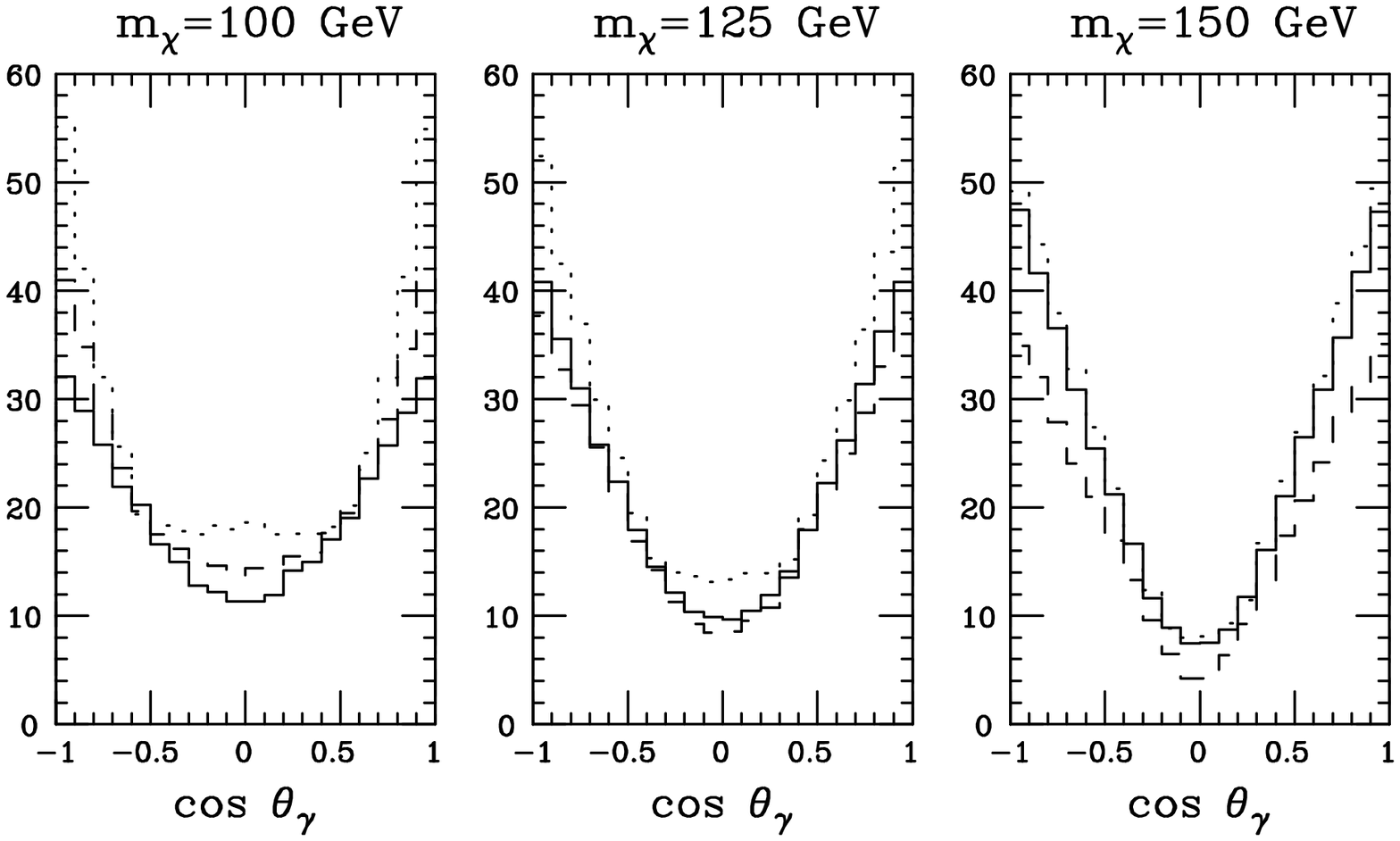}
\caption{Photonic angular distributions in neutralino(bino)-gravitino production at LEP190 for $m_\chi=100,125,150\,{\rm GeV}$ and $m_{\tilde e}=\rm 75$ (solid), 150 (dashed), and 300 (dots) GeV.}
\label{fig:Cbins}
\end{figure}

The observable photonic energy and angular distributions can be quite unwieldly
once we allow for the many choices of parameters that we have considered above.
An examination of all parameter combinations shows that both energy and angular distributions are largely insensitive to the neutralino composition, being
much more sensitive to the mass parameters ({\em i.e.}, $m_\chi,m_{\tilde e}$).
This result is perhaps not surprising as the observable distributions of relativistic particles are dominated by kinematical effects which depend
crucially on the mass parameters. Thus, for brevity we show only the result
in the bino case which, in any event, is representative of typical supergravity
models. The energy ($E_\gamma$) and angular ($\cos\theta_\gamma$) distributions
for $m_\chi=100,125,150\,{\rm GeV}$ are shown in Fig.~\ref{fig:Ebins} and
Fig.~\ref{fig:Cbins} respectively, for $m_{\tilde e}=75,150,300\,{\rm GeV}$.

The energy distributions (Fig.~\ref{fig:Ebins}) show a significant $m_\chi$
and $m_{\tilde e}$ dependence. As the neutralino mass grows, it tends to
produce harder photons. In fact, it is not hard to show that in the decay
$\chi\to\gamma\widetilde G$, with a neutralino energy and momentum as in
Eq.~(\ref{eq:Epchi}), the photon energy is restricted to the interval
\begin{equation}
{m^2_\chi\over2\sqrt{s}}<E_\gamma<{\sqrt{s}\over2}\ ,
\label{eq:Egamma}
\end{equation}
as faithfully reproduced in the simulations. (Near threshold ($m_\chi\approx\sqrt{s}$) the photon carries away half of the
center-of-mass energy.) These distributions show that any given single-photon candidate energy ($E_\gamma$) implies an upper bound on the possible
neutralino masses consistent with the candidate event,
\begin{equation}
m_\chi<\sqrt{2\sqrt{s}\,E_\gamma}\ .
\label{eq:mchimax}
\end{equation}

The photonic angular distributions (Fig.~\ref{fig:Cbins}) are peaked in the forward and backward directions, even more so as the neutralino becomes heavier. The selectron mass has an interesting effect. In the case of $m_\chi=100\,{\rm GeV}$, from Fig.~\ref{fig:Thetachi.bino} we see that for heavy selectron masses the neutralino angular distribution is fairly flat, and therefore the photonic distributions should reflect only kinematical effects, as they do ({\em i.e.}, peaked in the forward and backward directions). For light selectron masses, the neutralino avoids the forward and backward directions, and the kinematical effect on the photons is diminished. 

\subsection{Missing mass distribution}
The dominant background to the neutralino-gravitino signal is single radiative return to the $Z$:
$e^+e^-\to\gamma Z\to\gamma\nu\bar\nu$, where the photon is radiated off the
initial state and the $Z$ boson tends to be on shell. The most distinctive
signature for this background process appears in the missing invariant mass 
$M_{\rm miss}=\sqrt{E^2_{\rm miss}-p^2_{\rm miss}}$ distribution, which
is strongly peaked at $M_{\rm miss}\approx M_Z$. 

The missing mass distribution for the signal can be easily determined, as
in this case the missing energy and missing momentum are given by
\begin{equation}
E_{\rm miss}=\sqrt{s}-E_\gamma\,;\qquad 
p_{\rm miss}=|-\vec p_\gamma|=E_\gamma\ ,
\label{eq:Epmiss}
\end{equation}
and therefore
\begin{equation}
M_{\rm miss}=\sqrt{\sqrt{s}\,(\sqrt{s}-2E_\gamma)}\ .
\label{eq:Mmiss}
\end{equation}
The allowed range of $M_{\rm miss}$ is obtained by inserting the range of
photonic energies in Eq.~(\ref{eq:Egamma}); we obtain
\begin{equation}
0<M_{\rm miss}<\sqrt{s-m^2_\chi}\ .
\label{eq:Mmissrange}
\end{equation}
Histograms showing missing mass distributions fall in the range specified
by Eq.~(\ref{eq:Mmissrange}), and otherwise favor the upper end of the
$M_{\rm miss}$ range (corresponding to the lower end of the photonic energy
range). For brevity, we display these distributions only in the one-parameter
model example discussed in Sec.~\ref{sec:models} below. 

We note in passing that the complementary diphoton events from neutralino pair-production have a background ($e^+e^-\to\gamma\gamma Z\to\gamma\gamma\nu\bar\nu$) that also peaks for $M_{\rm miss}\approx M_Z$ \cite{DineKane}. In this case the missing mass in the diphoton signal varies from zero up to a maximum value of ${1\over2}(\sqrt{s}+\sqrt{s-4m^2_\chi})$, in contrast with the result for the single-photon signal in Eq.~(\ref{eq:Mmissrange}). The $M_{\rm miss}$ distributions of the single-photon and diphoton signals differ not only in range, but also in shape.

\section{One-parameter model example}
\label{sec:models}
It should have become clear from the discussion in Sec.~\ref{sec:signal},
that the signals to be searched for experimentally can have a wide range of
characteristics because of the variations in the underlying parameters
describing the neutralino-gravitino process. In reality, the model of
supersymmetry that describes nature will have all its mass parameters correlated in some way, and the actual observations may be a bewildering
composite of the many curves shown above. To exemplify this situation, 
in this section we specialize our results to the case of the one-parameter
no-scale supergravity model that has been mentioned at various places in
the preceding discussion. We have already shown the corresponding single-photon cross sections at LEP~1, LEP161, and LEP190, as the dashed lines in Figs.~\ref{fig:sigmaMz},\ref{fig:sigma161},\ref{fig:sigma190} respectively.
The cross sections for $\sqrt{s}>M_Z$ show a peculiar bump that, as discussed in Sec.~\ref{sec:LEP161}, can be traced back to the fact that the selectron masses vary with the neutralino mass.

\begin{figure}[p]
\vspace{5in}
\includegraphics{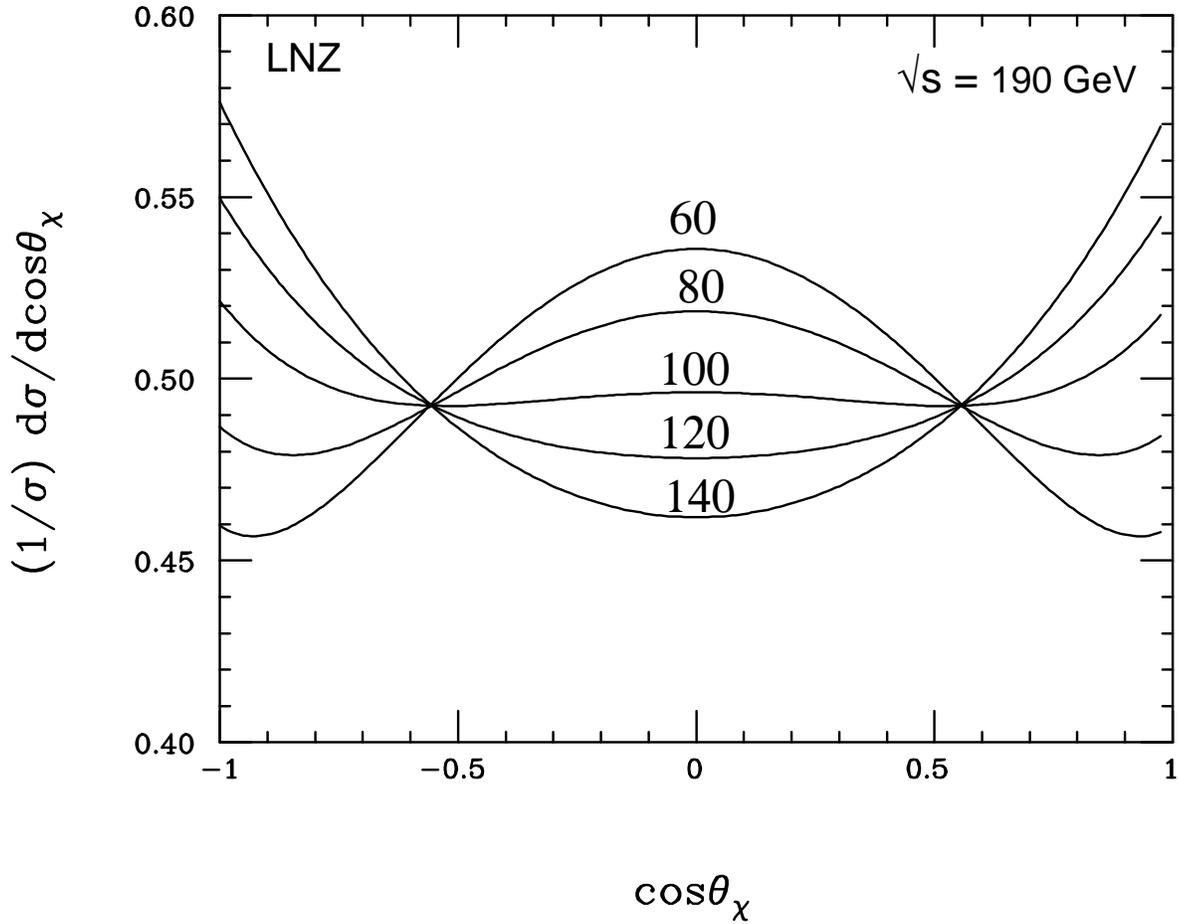}
\vspace{1cm}
\caption{Normalized angular distribution of neutralinos in neutralino-gravitino production at LEP190 for $m_\chi=60,80,100,120,140\,{\rm GeV}$ in a one-parameter no-scale supergravity model.}
\label{fig:ThetachiLNZ}
\end{figure}

As a first step towards obtaining the angular and energy photonic distributions, in Fig.~\ref{fig:ThetachiLNZ} we show the normalized 
neutralino angular distributions ${1\over\sigma}\,{d\sigma\over d\cos\theta_\chi}$ for $\sqrt{s}=190\,{\rm GeV}$ and
$m_\chi=60,80,100,120,140\,{\rm GeV}$. The total cross sections in each of these cases are $\sigma=1.2,1.3,1.0,0.6,0.2\,{\rm pb}$. Note how
relatively flat the angular distributions are: no more than a 10\% variation.
This is to be contrasted with the wide range of variability observed in
the generic cases shown in Figs.~\ref{fig:Thetachi.bino},
\ref{fig:Thetachi.zino}, and \ref{fig:Thetachi.photino}. In fact, the results
in the one-parameter model resemble those in the generic models when the
selectron mass is large ({\em i.e.}, the dotted lines in those figures,
which correspond to $m_{\tilde e}=300\,{\rm GeV}$). This is to be expected
as in the one-parameter model one has $m_{\tilde e_L}\sim 1.5\, m_{\tilde e_R}\sim2 m_\chi$, indicating increasingly heavier selectrons.

\begin{figure}[p]
\vspace{5in}
\includegraphics{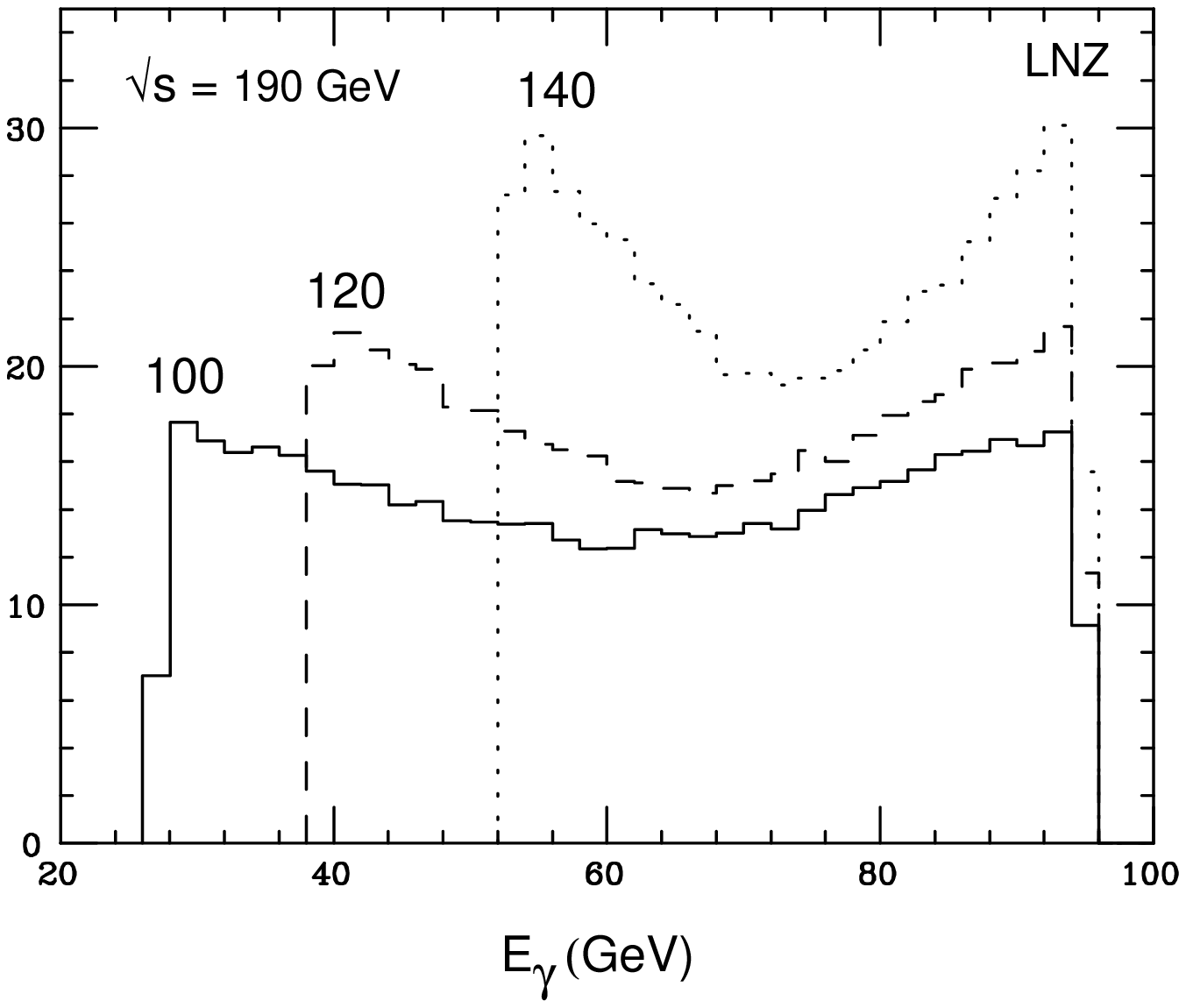}
\vspace{1cm}
\caption{Photonic energy distributions at LEP190 in a one-parameter no-scale supergravity model for $m_\chi=100,120,140\,{\rm GeV}$.}
\label{fig:EbinsLNZ}
\end{figure}

\begin{figure}[p]
\vspace{5in}
\includegraphics{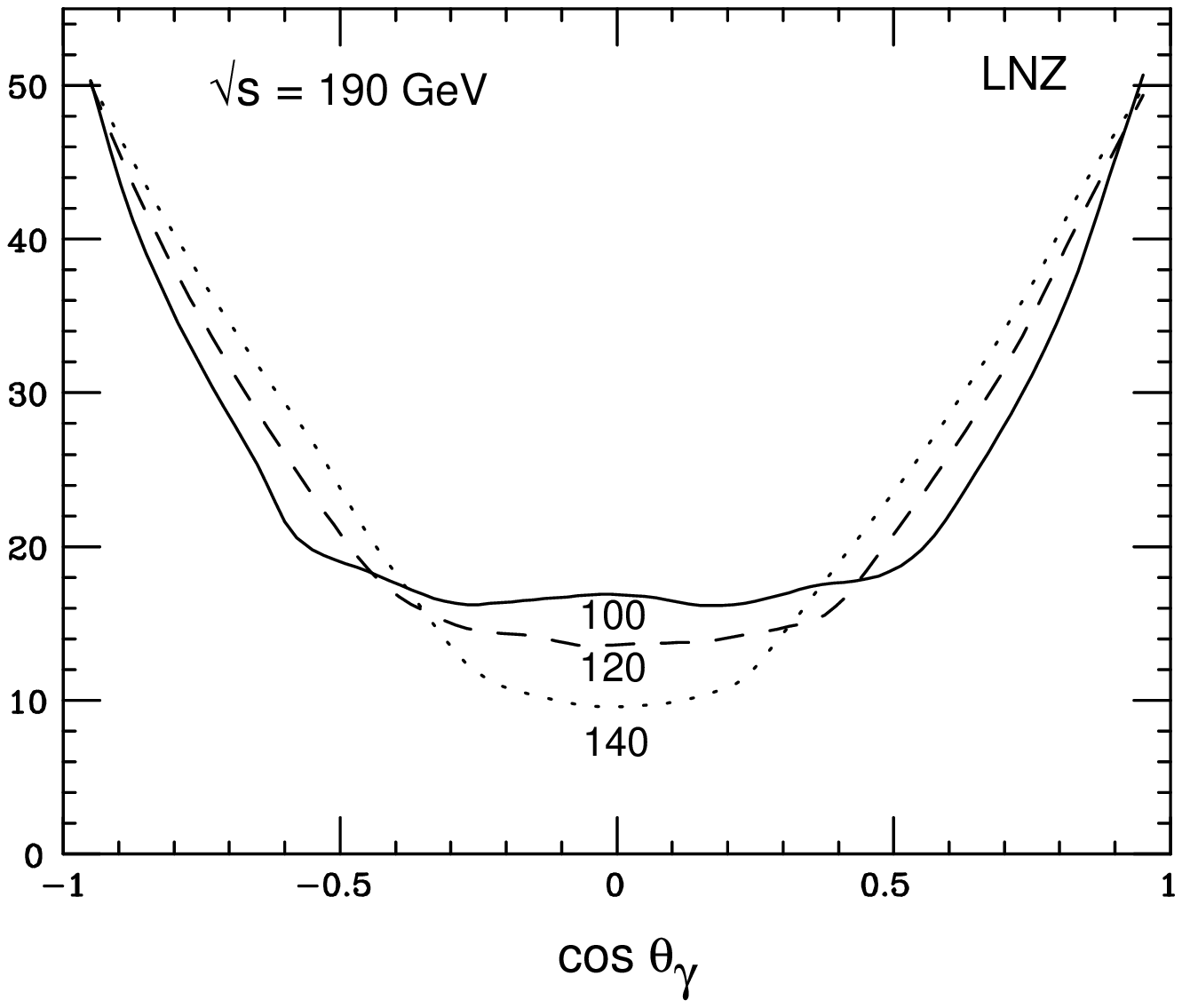}
\vspace{1cm}
\caption{Photonic angular distributions at LEP190 in a one-parameter no-scale supergravity model for $m_\chi=100,120,140\,{\rm GeV}$.}
\label{fig:CbinsLNZ}
\end{figure}

Following the method outlined in Sec.~\ref{sec:signal}, in Figs.~\ref{fig:EbinsLNZ} and \ref{fig:CbinsLNZ} we display the photonic energy
and angular distributions at LEP190 for three representative neutralino masses
($m_\chi=100,120,140\,{\rm GeV}$). The energy distributions show the same
restrictive photon energy behavior as predicted by Eq.~(\ref{eq:Egamma}).
The angular distributions are also peaked in the forward and backward
directions.

\begin{figure}[p]
\vspace{5in}
\includegraphics{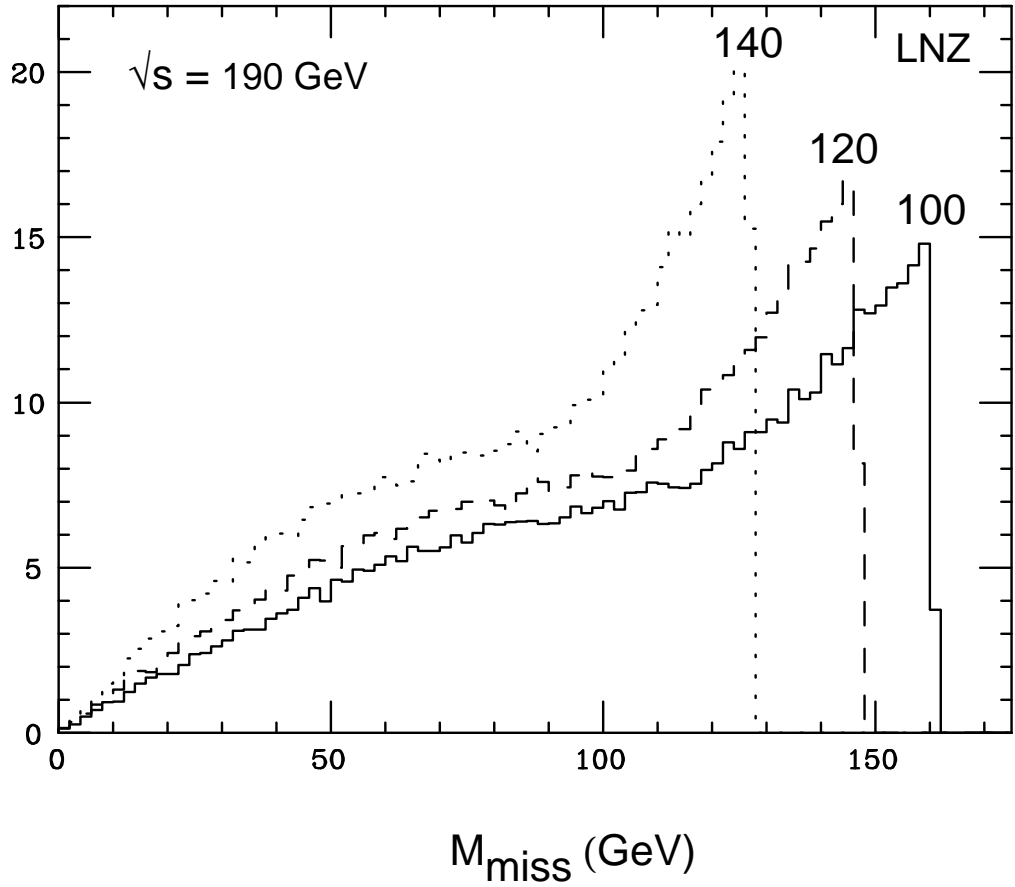}
\vspace{1cm}
\caption{Missing invariant mass distributions at LEP190 in a one-parameter no-scale supergravity model for $m_\chi=100,120,140\,{\rm GeV}$.}
\label{fig:MbinsLNZ}
\end{figure}

Finally we consider the missing mass distributions, which are obtained from
Eq.~(\ref{eq:Mmiss}), and are shown in Fig.~\ref{fig:MbinsLNZ}. We note the
range of $M_{\rm miss}$, as prescribed by Eq.~(\ref{eq:Mmissrange}), and
the tendency to favor missing mass values toward the upper end of the allowed
interval. For the neutralino mass choices shown, the missing mass shows a
distinct preference to be larger than $M_Z$. (This is in contrast with the
$M_{\rm miss}$ distribution in diphoton events, which is more evenly distributed.) 

\section{Conclusions}
\label{sec:conclusions}
In this paper we have attempted to study in some detail the physics of
supersymmetric models with a gravitino light enough that it can be
produced directly at collider experiments. Our discussion has centered
mainly around LEP, from where the strongest constraints can be obtained
at the moment. We have nonetheless outlined the corresponding program to
be followed at the Tevatron, where instrumental backgrounds make identification
of the single-photon signal a more challenging task.

We have provided new and explicitly analytical expressions for neutralino-gravitino differential cross sections at $e^+e^-$ colliders
and have discussed some of the theoretical subtleties involved in the calculation and some of the peculiar parameter dependences of the cross section. We have used our expressions to obtain new lower bounds from LEP~1
data on the gravitino mass for $m_\chi<M_Z$. Weaker limits from LEP~2 are obtained at higher neutralino masses. Our study includes a Monte Carlo simulation of the single-photon signal, which should be helpful in the experimental analyses that are just now getting underway. We have also
specialized the results to our one-parameter no-scale supergravity model,
where the signals can be analyzed much more simply because of the tight
correlations between the model parameters.

\section*{Acknowledgments}
J.~L. would like to thank G. Eppley, T. Gherghetta, and T. Moroi for useful discussions. The work of J.~L. has been supported in part by DOE grant DE-FG05-93-ER-40717 and that of D.V.N. by DOE grant DE-FG05-91-ER-40633.

\clearpage

\clearpage

\clearpage

\clearpage

\clearpage

\clearpage

\clearpage

\clearpage

\clearpage

\clearpage

\clearpage

\clearpage

\clearpage

\clearpage

\end{document}